\documentclass[useAMS,usenatbib]{mn2e}
\usepackage{graphicx, amsmath, amssymb,epsfig}
\usepackage{mn2e-breakabs}
\usepackage{graphicx}
\usepackage{subfigure}
\usepackage{times}
\usepackage{caption}
\usepackage{rotating}
\usepackage{rotfloat}
\usepackage{multicol}
\usepackage{array}
\usepackage{booktabs}

\def\kpch{\mbox{$h^{-1}$kpc}}

\def\LCDM{\mbox{$\Lambda$CDM}}

\def\Mpch{\mbox{$h^{-1}$Mpc}}
\def\Mpcvol{\mbox{$h^{3}$Mpc$^{-3}$}}
\def\Gpcvol{\mbox{($h^{-1}$Gpc)$^{3}$}}
\def\Gpch{\mbox{$h^{-1}$Gpc}}

\def\M200{\mbox{$M_{\rm 200}$}}

\def\Msunh{\mbox{$h^{-1}M_\odot$}}

\def\R200{\mbox{$R_{\rm 200}$}}

\def\V200{\mbox{$V_{\rm 200}$}}

\newcommand{\mnras}{MNRAS}

\newcommand{\mpcoh}{\,h^{-1}\,{\rm Mpc}}

\def\la{\mathrel{\mathpalette\fun <}}
\def\ga{\mathrel{\mathpalette\fun >}}
\def\fun#1#2{\lower3.6pt\vbox{\baselineskip0pt\lineskip.9pt
        \ialign{$\mathsurround=0pt#1\hfill##\hfil$\crcr#2\crcr\sim\crcr}}}

\newcommand{\be}{\begin{equation}}
\newcommand{\ee}{\end{equation}}
\newcommand{\ba}{\begin{eqnarray}}
\newcommand{\ea}{\end{eqnarray}}
\newcommand{\simgt}{\,\hbox{\lower0.6ex\hbox{$\sim$}\llap{\raise0.6ex\hbox{$>$}}}\,}
\newcommand{\simlt}{\,\hbox{\lower0.6ex\hbox{$\sim$}\llap{\raise0.6ex\hbox{$<$}}}\,}

\begin{document}

\title[Hunting down systematics in BAO]
{Hunting down systematics in baryon acoustic oscillations after cosmic high noon}

\author[Prada et al.]{
  \parbox{\textwidth}{
Francisco Prada$^{1,2,3}$\thanks{E-mail: f.prada@csic.es}, Claudia G. Sc\'occola$^{1,4,5,6}$, Chia-Hsun Chuang$^{1}$\thanks{MultiDark Fellow}, Gustavo Yepes$^{4}$, \\
Anatoly A. Klypin$^{7}$, Francisco-Shu Kitaura$^{8}$, Stefan Gottl\"{o}ber$^{8}$, and Cheng Zhao$^{9}$
 }
  \vspace*{4pt} \\
$^1$ Instituto de F\'{\i}sica Te\'orica, (UAM/CSIC), Universidad Aut\'onoma de Madrid,  Cantoblanco, E-28049 Madrid, Spain \\
$^2$ Campus of International Excellence UAM+CSIC, Cantoblanco, E-28049 Madrid, Spain \\
$^3$ Instituto de Astrof\'{\i}sica de Andaluc\'{\i}a (CSIC), Glorieta de la Astronom\'{\i}a, E-18080 Granada, Spain \\
$^4$ Departamento de F\'{\i}sica Te\'orica, Universidad Aut\'onoma de Madrid, Cantoblanco, 28049, Madrid, Spain \\
$^{5}$ Instituto de Astrof{\'\i}sica de Canarias (IAC), C/V{\'\i}a L\'actea, s/n, La Laguna, Tenerife, Spain. \\
$^{6}$ Dpto. Astrof{\'\i}sica, Universidad de La Laguna (ULL), E-38206 La Laguna, Tenerife, Spain.\\
$^7$ Astronomy Department, New Mexico State University, Las Cruces, NM 88003, USA\\
$^8$ Leibniz-Institut fuer Astrophysik (AIP), An der Sternwarte 16, D-14482 Potsdam, Germany\\
$^9$ Tsinghua Center for Astrophysics, Department of Physics, Tsinghua University, Haidian District, Beijing 100084, P. R. China \\
}

\date{\today} 

\maketitle

\begin{abstract}
Future dark energy experiments will require better and more accurate theoretical predictions for the baryonic acoustic oscillations (BAO) signature in the spectrum of cosmological perturbations. Here, we use large $N$-body simulations of the $\Lambda$CDM Planck cosmology to study any possible systematic shifts and damping in BAO due to the impact of nonlinear gravitational growth of structure, scale dependent and non-local bias, and redshift-space distortions.  The effect of cosmic variance is largely reduced by dividing the tracer power spectrum by
that from a ”BAO-free” simulation starting with the same phases. The high accuracy of our simulations allows us to resolve well dark matter halos and subhalos inside them. This is crucial to obtain robust results, as opposed to the majority of previous studies which due to the
lack of resolution  have to rely on statistical prescriptions to connect dark matter with galaxies. This permits us to study with unprecedented accuracy (better than $0.02\%$  for dark matter and $0.07\%$ for low-bias halos) small shifts $\alpha$ of the pristine BAO wavenumbers towards larger $k$, and non-linear damping $\Sigma_{\rm nl}$ of BAO wiggles in the power spectrum of dark matter and halo populations in the redshift range $z=0-1$. For dark matter, we provide an accurate parametrization of the evolution of $\alpha$ as a function of the linear growth factor $D(z)$. For halo samples, with bias ranging from $1.2$ to $2.8$, we measure a typical BAO shift of $\approx 0.25\%$, observed in real-space, which does not  show an appreciable evolution with redshift within the uncertainties. Moreover, we report a constant shift as a function of halo bias. We find a different evolution of the damping of the acoustic feature in all halo samples as compared to dark matter with haloes suffering less damping, and also find some weak dependence on bias. A larger BAO shift and damping is measured in redshift-space which can be well explained by linear theory due to redshift-space distortions.  A clear modulation in phase with the acoustic scale is observed in the scale-dependent halo bias due to the presence of the baryonic acoustic oscillations.  The work presented in this paper settles an optimal strategy for studying the nonlinear BAO systematics with high performance $N$-body simulations and contributes towards a deeper theoretical understanding for upcoming large BAO surveys.

\end{abstract}

\begin{keywords}
  (cosmology:)  large-scale structure of
  Universe - galaxies: haloes - galaxies: statistics - dark matter
\end{keywords}

\section{Introduction}
The discovery of cosmic acceleration has motivated the development of large experiments that aim at measuring the expansion history of the universe and growth of structure with high precision at the $0.1-1\%$ level. More precise measurements of the Baryon Acoustic Oscillation (BAO) scale rely on ongoing and future large galaxy, QSO and Ly$\alpha$ surveys and improved analysis techniques. This field has undergone an enormous progress since the BAO peak was detected for the first time in the SDSS-II and 2dFGRS galaxy clustering statistics \citep{Eisenstein:2005su,Cole2005}.  The SDSS-III/BOSS survey, with almost four years of data, has already reached $1.0\%$ precision on measuring the baryon acoustic scale using the DR11 CMASS sample of massive galaxies at $z=0.57$ \citep{Anderson2014}. This is a significant achievement compared to the  $4\%$ precision of the first SDSS-II LRG measurements. It is worth mentioning the BAO measurements that have been conducted at $z\sim0.35$ using the SDSS-II DR7 LRG  \citep[e.g.][]{Percival:2007yw,Sanchez:2009jq,Reid:2010xm,Chuang:2012dv,Padmanabhan:2012hf} and WiggleZ \citep{Blake:2011wn} $z\sim0.6$ survey data.  

With the completion of BOSS after DR12, the $1\%$ precision will be superseded by new experiments such as DESI \citep{Schelgel:2011zz,Levi2013}  and Euclid \citep[e.g.][]{Laureijs:2011}. Both surveys aim at measuring the BAO scale to the sub-percent level over a wide redshift range $0.5 < z < 3.5$, thus, providing unprecedented constraints on the dark energy equation of state (see \citet{Weinberg2012} for a complete review and forecasts on cosmological models with current and future planned BAO experiments). The new generation dark energy experiments also impose severe challenges on understanding any possible systematic shifts in the BAO signature due to nonlinear gravitational growth, scale-dependent bias and redshift space distortions (RSD) to a high precision, better than the measured statistical uncertainties.  This challenge has motivated in the recent years many works based on perturbation theory and large-volume $N$-body simulations to understand the damping and shifts of the BAO feature as being probe by dark matter and biased tracers \citep[e.g.][]{Angulo2008,Seo2008,Smith2008,Sanchez2008,CroSco2008,Nikhil2009,Seo2010,Mehta2011,Zaldarriaga2012, Angulo2013,Rasera2014}.

A shift in the acoustic scale of $\alpha-1\sim0.3 [\%]$  has been measured in the dark matter power spectrum at $z=0$ using $N$-body simulations, where $\alpha$ is the ratio of the linear BAO scale to the measured scale. In this case, the BAO feature is found toward larger $k$, relative to the linear P(k) \citep[see e.g.][]{Seo2010}. This shift, and its dependence with redshift, has been well explained by perturbation theory in numerous works as due to additional oscillations generated by non-linear mode coupling effects  \citep[e.g.][]{CroSco2008,Nikhil2009,Zaldarriaga2012}.  Yet, the situation with dark matter halos is not that clear due to the lack of mass and force resolution of the simulations used for the measurements \citep[see e.g.][]{Angulo2008,Nikhil2009,Mehta2011}. For example, \citet{Mehta2011} did not detect any shift in the acoustic scale for their halo models with $b<3$ biased tracers. On the other hand, perturbation theory does not provide a solid prediction for these shifts in the halo clustering statistics. The shift seems to depend on two halo bias parameters, $b_1$ and $b_2$, which in principle will cause possible arbitrary shifts of the acoustic scale \citep[see][]{Nikhil2009}.

In this work, we investigate and measure the nonlinear shift of the acoustic scale in the dark matter halo power spectrum relative to the underlying dark matter distribution taking advantage of the new suite of BigMultiDark simulations (hereafter BigMD) which combines high-resolution with large-volume for the adopted $\LCDM$ standard cosmological model. The BigMD simulations are designed to have sufficient resolution to resolve halos and subhalos within a cubic box of $2.5 \Gpch$ on a side with a completeness suitable to study the clustering of galaxies hosted by halo samples with bias down to 1.2 at $z=0$. This bias regime has never been explored with decent numerical resolution for the study of the non-linear BAO shift to a high-precision. For consistency, we have also studied the damping of acoustic oscillations both for dark matter and halo biased tracers.

The paper is organised as follows. In Sec. \ref{sec:sim}, we introduce the BigMD simulation data used in our study. In Sec. \ref{sec:bao}, we describe the details of the methodology adopted to measure the BAO shift and damping for dark matter and various halo samples. In Sec. \ref{sec:results}, we present our main results, and  
we summarise and conclude in Sec.~\ref{sec:conclusion}.


\begin{figure*}
\begin{tabular}{cc}
\includegraphics[width=9.1cm]{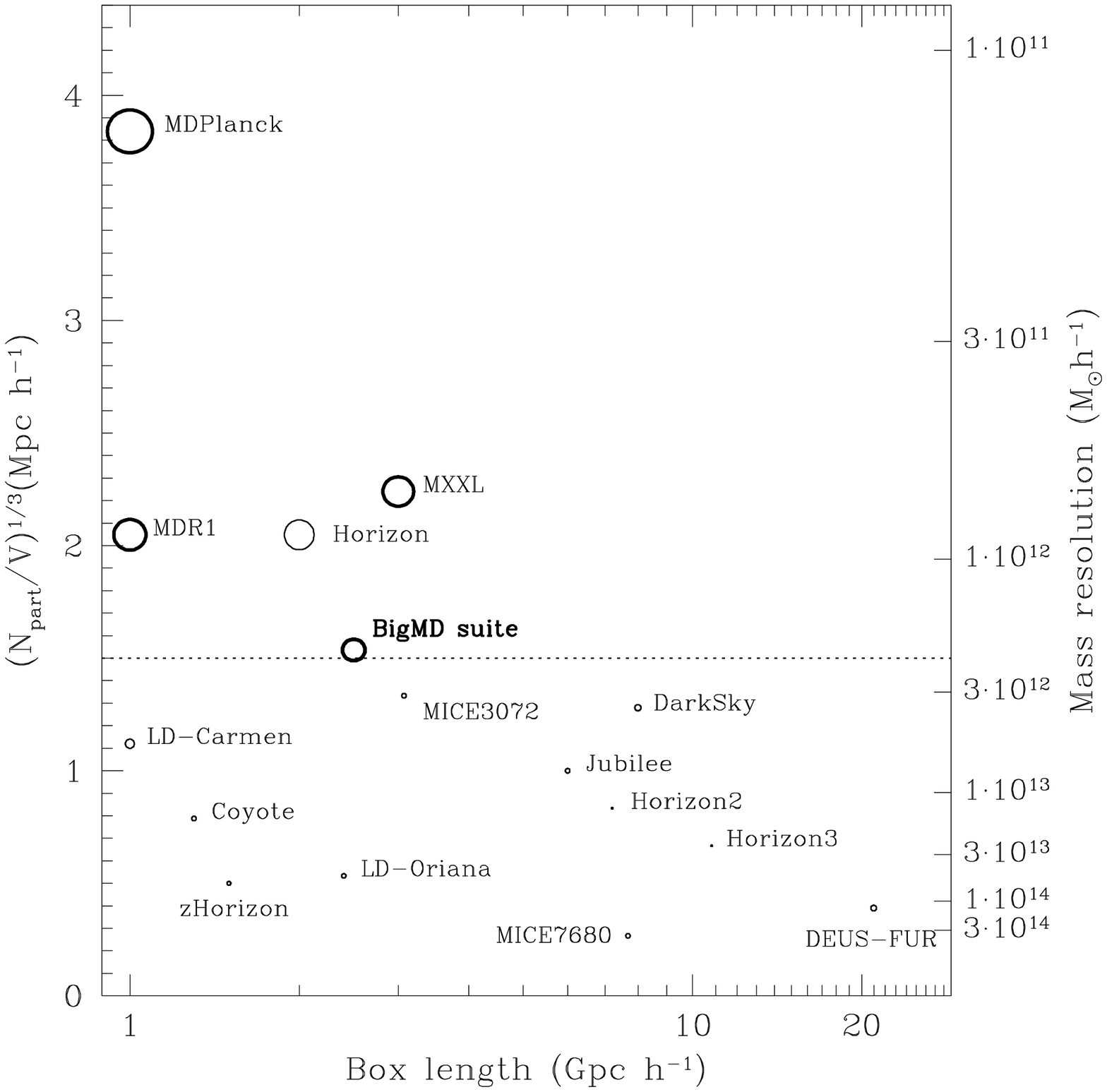}
\includegraphics[width=9.1cm]{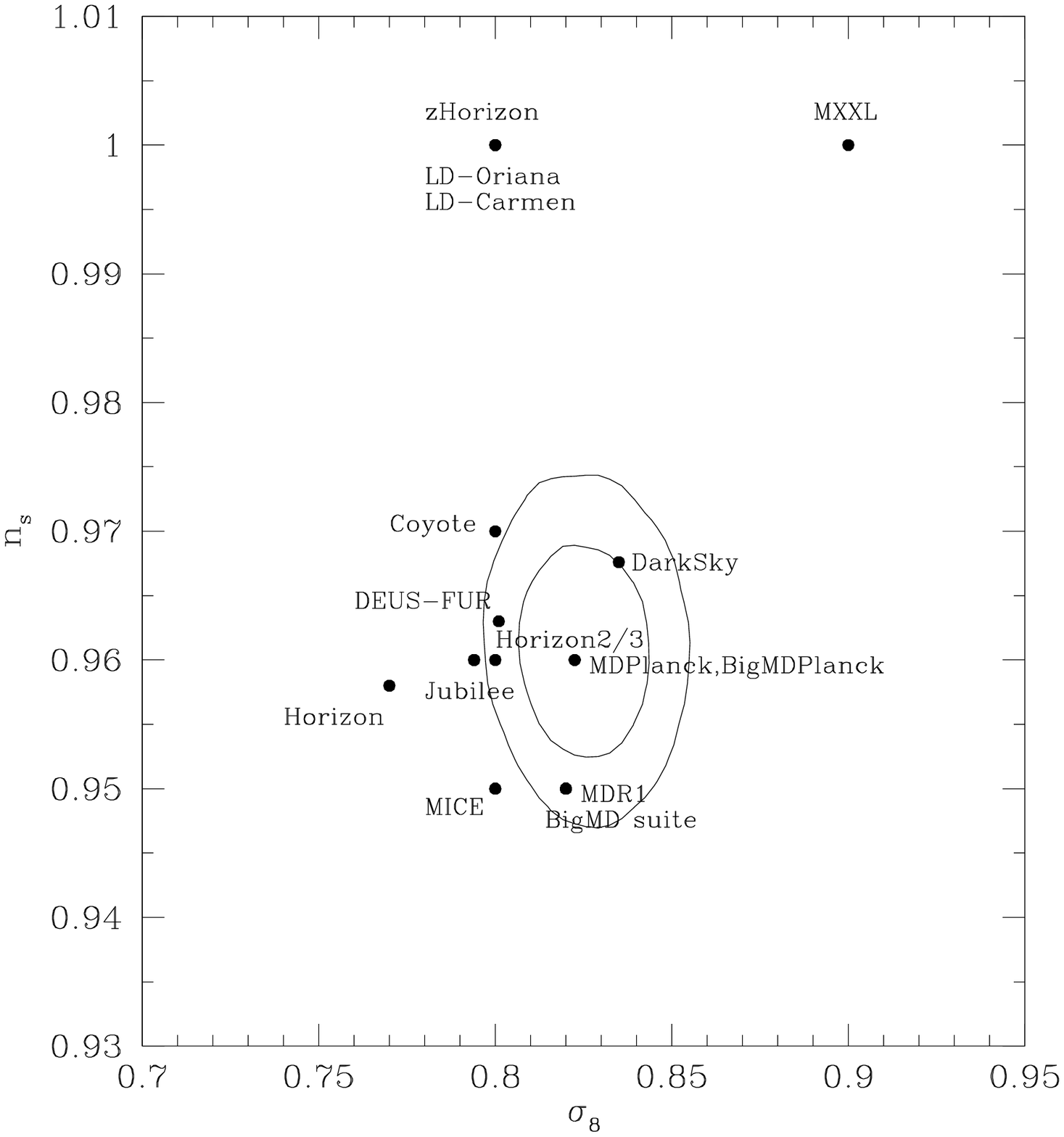}
\end{tabular}
\caption{\label{fig:comsim} Left: Compilation of the basic numerical parameters adopted in large $N$-body cosmological simulations used in recent years for galaxy clustering and bias studies. The number of particles per unit comoving distance (and mass resolution for halos with at least 100 particles) is shown as a function of the box length for each simulation. The size of the circles is inversely proportional to the softening parameter $\epsilon$ used in the gravitational force. Our new suite of BigMultiDark simulations have been designed to meet all science requirements needed to interpret the galaxy clustering in the BOSS survey (dotted line). Right: $n_s$ versus $\sigma_8$ (right) cosmological parameters adopted for each simulation. Contours show $68\%$ and $95\%$ confidence levels from Planck assuming a flat $\Lambda$CDM Planck cosmology. In this work we are using the BigMD Planck simulations.}
\end{figure*}

\section{Simulations for large galaxy surveys} 
\label{sec:sim}

Figure~\ref{fig:comsim} displays an overview of the basic numerical (force and mass resolution) and cosmological parameters adopted in state-of-the-art cosmological simulations, comprising at least one $\Gpcvol$ in volume, carried out to study galaxy clustering and bias for large galaxy surveys \citep[i. e. Horizon, MICE7680/MICE3072, LD-Carmen/LD-Oriana, Horizon2/Horizon3, DEUS-FUR, MXXL, MDR1, zHorizon, BigMD-suite/MDPL, Coyote, Jubilee, and DarkSky by][respectively]{Teyssier2009,Crocce2010,McBride,Kim2011,Alimi2012,Angulo2012,Prada2012,Smith2012,Klypin2014,Lawrence2010,Watson2013,Skillman2014}. In the left panel, for each simulation box we plot the number of particles per unit comoving distance (and the mass resolution for halos with at least 100 particles) versus the simulation box length. The size of the circles are inversely proportional to the softening parameters, $\epsilon$, used in the gravitational force: the larger circle corresponds to MultiDark Planck (MDPL) with $\epsilon=5\kpch$ and the smallest to the Horizon-3 run with $\epsilon=150\kpch$.  We show, in the right panel, some of the key cosmological parameters assumed in each simulation, $n_s$ (the spectral index of the primordial power spectrum) and $\sigma_8$ (the $rms$ amplitude of linear mass fluctuations in spheres of $8\Mpch$ comoving radius at redshift $z=0$), compared with the Planck $68\%$ and $95\%$ confidence level contours assuming a flat $\LCDM$ cosmology \citep[][]{Planck2013}.

The BigMD-suite of $\LCDM$ simulations have been designed to meet the science requirements of the BOSS galaxy survey, i.e. the numerical requirements for mass and force resolution that allows to resolve well those halos and subhalos that can host typical BOSS massive galaxies at $z\sim0.5$, which will permit to create mock catalogs with the appropriate galaxy bias and clustering. The baseline of the BigMD $N$-body simulations comprises $3840^3$ particles in a box with $2.5 \Gpch$ on a side.  Initial conditions were set at the redshift $z_{init}=100$ with identical Gaussian fluctuations for all simulations. We used GINNUNGAGAP\footnote{ http://code.google.com/p/ginnungagap}, a publicly available full MPI+OpenMP initial conditions generator  code that  uses  Zeldovich Approximation with an unlimited number of particles. The BigMD simulations were run with the L-GADGET-2 code \citep[see][for details]{Klypin2014}. In this work, we use a couple of those simulations where we adopted the cosmological parameters based on the latest fits to the Planck data  \citep[][]{Planck2013}. The mass and force resolutions are $2.36\times10^{10}\Msunh$ and $10\kpch$. The choice of numerical parameters to meet our requirements (combination of mass and force resolutions is highlighted with a dashed-line in Figure~\ref{fig:comsim}) were chosen after the completion of many tests to study the convergence for the correlation function and circular velocities for halos and their subhalos \citep[see][for details]{Klypin2013}. This allows us to resolve well the internal structure of (sub)halos, thus, making possible to connect them with BOSS-like galaxies. 
 
\begin{figure}
\begin{tabular}{cc}
\includegraphics[width=8.cm]{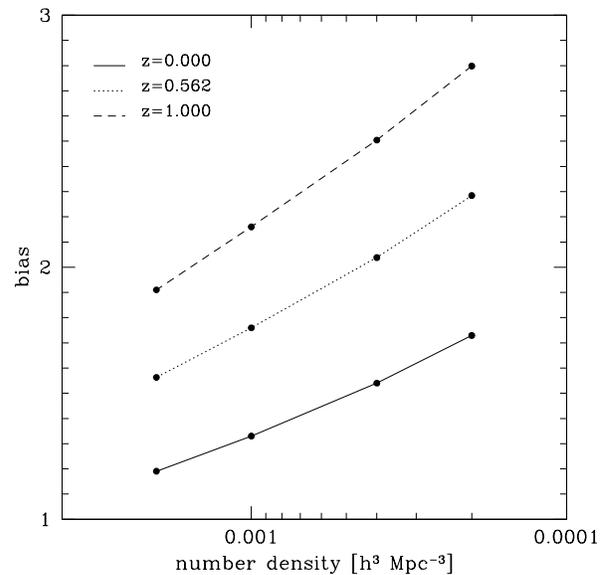}
\end{tabular}
\caption{\label{fig:bias} Bias as a function of number density at $z=0, 0.562$, and $1$, for four different samples of dark matter halos selected accordingly to their maximum circular velocities $V_{max}$ from our BigMD Planck simulations.}
\end{figure}

  Dark matter halos (and subhalos) were identified with a parallel version of the Bound-Density-Maxima (BDM) algorithm \citep[][]{KH1997,Riebe2011}. BDM is a spherical overdensity code that provides many properties of halos and subhalos in our BigMD simulations. We then use a simple, non-parametric Halo Abundance Matching (HAM) prescription, to connect dark matter (sub)halos  with galaxies by selecting them above a given maximum circular velocity $V_{max}$. This procedure is able to predict the clustering properties, and the halo occupation distribution of observed galaxies for different number densities \citep[e.g.][]{Conroy2006,T2011,Nuza:2012mw}. We selected four different halo samples from the BigMD BDM catalogs for our analysis with number densities $2\times10^{-3}, 1\times10^{-3}, 4\times10^{-4}, 2\times10^{-4} \Mpcvol$,  corresponding to linear biases 1.56, 1.76, 2.04, and 2.28 at several redshifts up to $z=1$ (see  Figure~\ref{fig:bias}), of typical Emission Line Galaxies (ELG) and Luminous Red Galaxies (LRGs) as those being targeted in the major surveys discussed here.
  
\section{Measuring the shift and damping of the acoustic scale}
\label{sec:bao}

The analysis performed in this work on the nonlinear evolution of the shift and damping of acoustic oscillations in the power spectrum of dark matter and the halo samples mentioned above, is based on two simulations of our BigMD suite. The first one, BigMDPL, adopted the initial matter power spectrum generated using CAMB \citep{Lewis:1999bs} with the Planck cosmological parameters;  and for the second BigMDPLnw simulation, the same cosmological parameters were assumed but with a smooth initial power spectrum with no BAO wiggles, generated by fitting a cubic spline \citep{press92} to the CAMB table with three nodes fixed empirically. We recall that the initial conditions of both simulations were generated with the same Gaussian amplitude and phases. Hence, the effect of cosmic variance is greatly reduced when dividing the  spectrum $P(k)$ computed from BigMDPL with BAOs, for a given tracer, by the non-wiggle BigMDPLnw power spectrum $P_{nw}(k)$. This allow us to obtain measurements with unprecedented accuracy of the BAO stretch parameter $\alpha$ and damping as a function of redshift. Below, we provide more details on the uncertainty in the BAO shift measurements.

We compute P(k) using  285 linear bins, in the $k$-range (0.085,0.8). The limits are chosen to contain the BAO oscillations, and have large enough range at smaller scales where BAO have effectively vanish out, which allows to estimate the errors. We use a Fourier mesh of  2500$^3$ cells  in a box size of (2.5  $h^{-1}$ Gpc)$^3$.  Density fields are calculated using the Cloud-In-Cell assignment scheme, and  aliasing and shot noise corrections were applied. To improve the convergence of the fitting, in redshift space, and only for the halos, we reduced the $k$-range to (0.085,0.6), and therefore the number of linear bins used was 205.

\begin{figure*}
\begin{tabular}{cc}
\includegraphics[width=8.8cm]{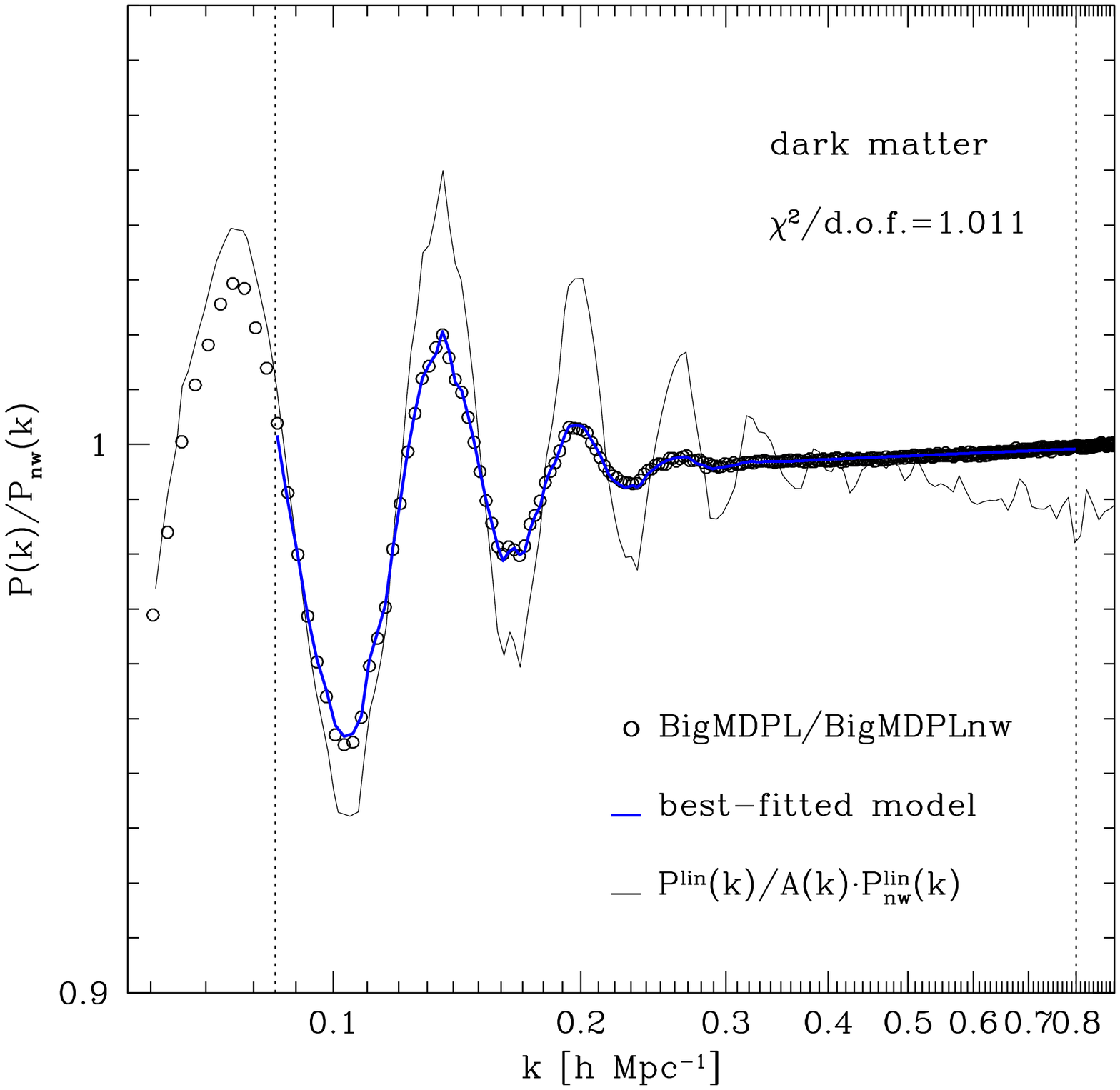}
\includegraphics[width=8.8cm]{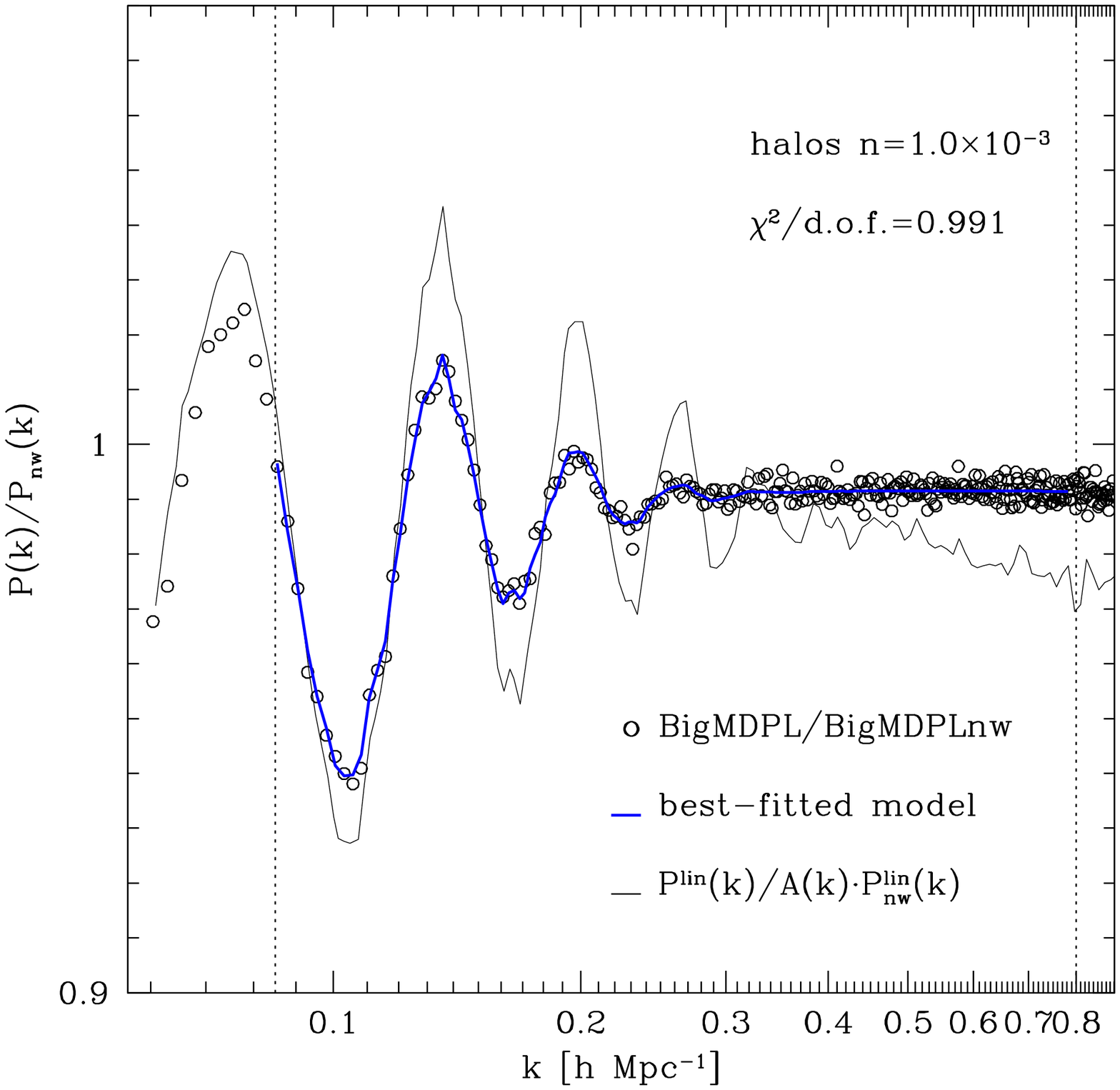} \\
\includegraphics[width=8.8cm]{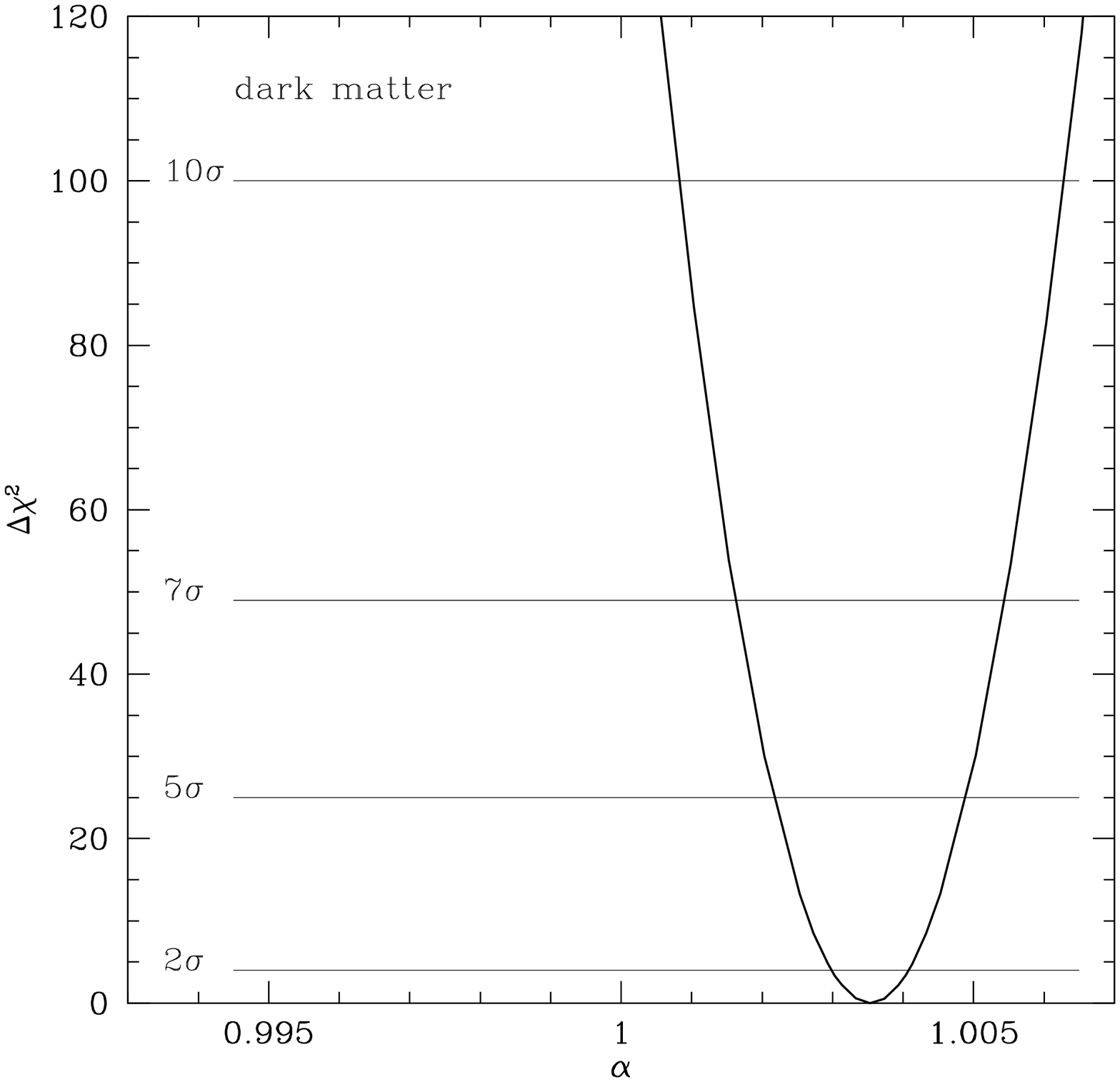}
\includegraphics[width=8.8cm]{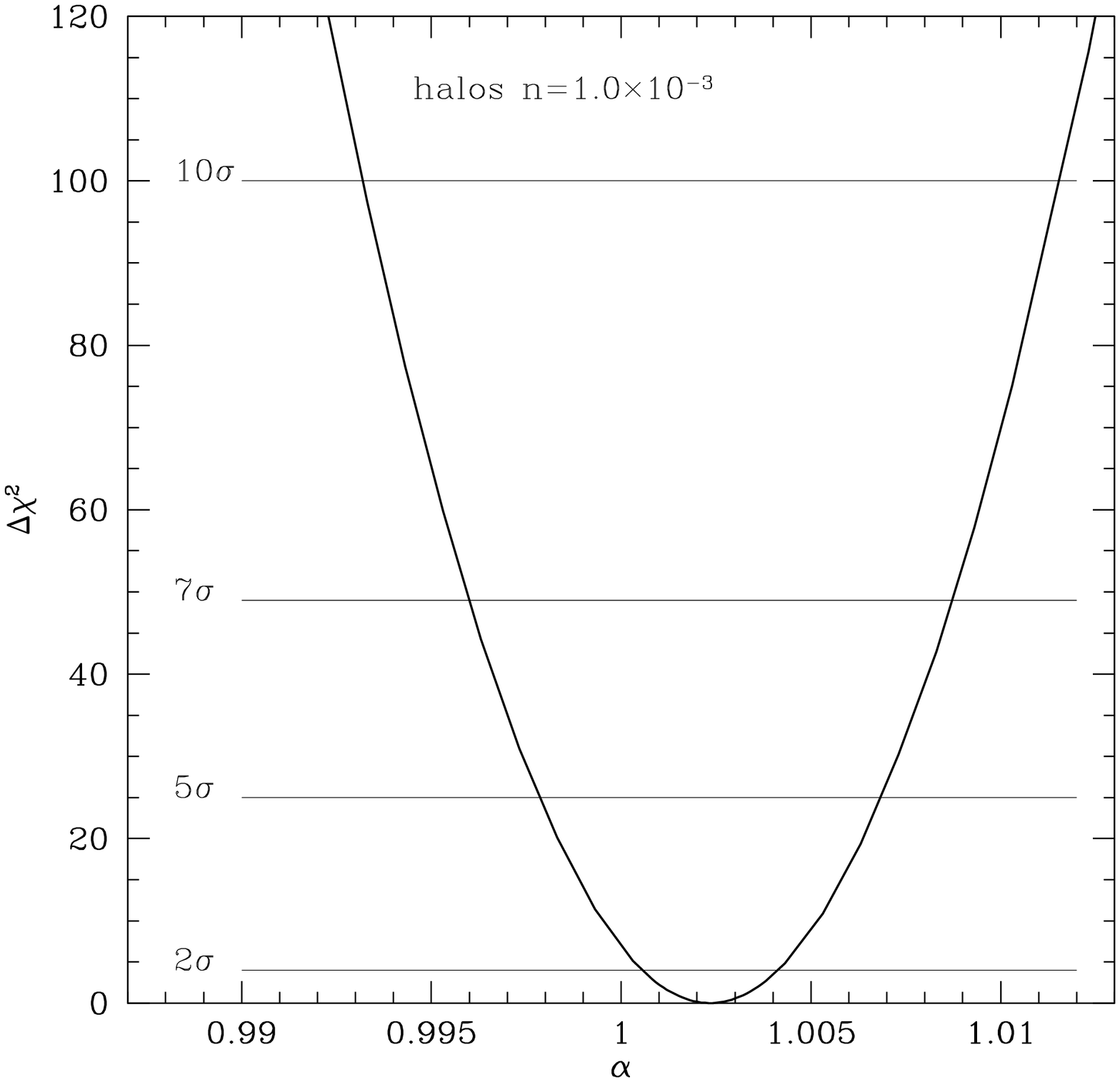}
\end{tabular}
\caption{\label{fig:power} Power spectra at $z=0$, in real-space, divided by the corresponding non-wiggle power spectrum obtained from BigMDPL and BigMDPLnw respectively for dark matter (top left panel) and a typical halo sample with number density 1 $\times 10^{-3}$ Mpc$^{-3}$ h$^3$ (top right panel). The thick solid line corresponds to the best-fitted model given by Eq.~\ref{eq:pkratio} in the wavenumber range $0.085  < k < 0.8$ h Mpc$^{-1}$ shown by the vertical dotted lines. The ratio of wiggle and non-wiggle linear matter power spectrum is also shown in both cases (thin solid line). The bottom panels show the likelihood $\chi^2$ distributions for the BAO shift $\alpha$ parameter both for dark matter (left panel) and the halo sample (right panel).}
\end{figure*}

Figure~\ref{fig:power} shows the spherically-average power spectra at $z=0$ in real-space drawn from BigMDPL, divided by the corresponding non-wiggle BigMDPLnw power spectrum for dark matter (top left panel) and a typical halo sample with number density 1 $\times 10^{-3} Mpc^{-3} h^3$ and $bias=1.33$ (top right panel). We measure the shift of the BAO relative to linear theory by following a similar methodology as that presented in \citet[][]{Seo2008}. For a given tracer, the power spectrum with wiggles is modeled by damping the acoustic oscillation features of the linear power spectrum assuming a Gaussian with a scale parameter $\Sigma_{nl}$ which accounts for the BAO broadening due to nonlinear effects \citep[e.g.][]{Eisenstein:2006nj}, i.e.

\begin{equation} \label{eq:eq1}
\begin{split}
P(k)=[(P^{lin}(k) - A(k)  P^{lin}_{nw}(k))  \exp\left(-k^2 \Sigma_{nl}^2/2\right) \\
+A(k)  P^{lin}_{nw}(k)] B(k) \, , 
\end{split}
\end{equation}
where $P^{lin}$ is the linear power spectrum generated with CAMB, adopting the Planck cosmology, and $P^{lin}_{nw}$ is the smooth non-wiggle spline power spectrum. $B(k)$ represents the non-linear growth of the matter power spectrum, which in the case of halos includes also a scale-dependent bias. The $A(k)$ term allows for any correction that might be needed to account for the proper description of the broad-band shape of the power spectrum. Note that $A(k)=1$ for an ideal case.
 
We then fit the ratio $P/P_{nw}$ of the power spectrum with acoustic oscillations to that with no-BAO drawn from the BigMDPL/BigMDPLnw simulation pair (see Figure~\ref{fig:power}) with the following formula, 

\begin{equation} \label{eq:pkratio}
\begin{split}
P(k)/P_{nw}(k)=\left[\left(\frac{P^{lin}(k/\alpha)}{A(k) P^{lin}_{nw}(k/\alpha)}-1\right)  \exp\left(-k^2 \Sigma_{nl}^2/2\right)\right. \\
 +\left. 1\right] C(k),
\end{split}
\end{equation}
where $C(k)$ accounts for the non-linear growth of both wiggle $P(k)$ and non-wiggle $P_{nw}(k)$ power spectra. Similar to \citet{Anderson2014}  we adopt simple power-law polynomials for both $A(k)$ and $C(k)$ terms expressed in the form $a_0 k^{a_1}$ and $c_0 k^{c_1}$ respectively. The shift and damping of the acoustic oscillations, measured by $\alpha$ and $\Sigma_{nl}$, are considered free parameters in our fit. For the $\chi^2$ analysis, we have 6 fitting parameters $\{ \alpha, \Sigma_{nl}, a_0, a_1, c_0, c_1\}$, and the fit is performed over the wavenumber range $0.085  < k < 0.8$ h Mpc$^{-1}$ (vertical dotted lines in Figure~\ref{fig:power}). Note that we avoid in the fit the first acoustic peak being distorted, up to some extend, by our own choice of the broad-band shape of the power spectrum when we build the featureless ("BAO free") power spectrum $P_{nw}(k)$. The $\chi^2$ per degree-of-freedom $\chi^2/d.o.f.$, which indicates the goodness-of-fit between our model and the BigMD simulation data, is 1.011 and 0.911 for the ratio of the wiggle to non-wiggle power spectra for dark matter and the halo sample shown in Figure~\ref{fig:power}. The solid line corresponds to the best-fitted model given by Eq.~\ref{eq:pkratio}. The damping of the BAO features is clearly seen in both dark matter and halos when compared with the linear wiggle to non-wiggle $P(k)$ ratio (thin solid line).

\begin{figure}
\begin{tabular}{cc}
\includegraphics[width=8.cm]{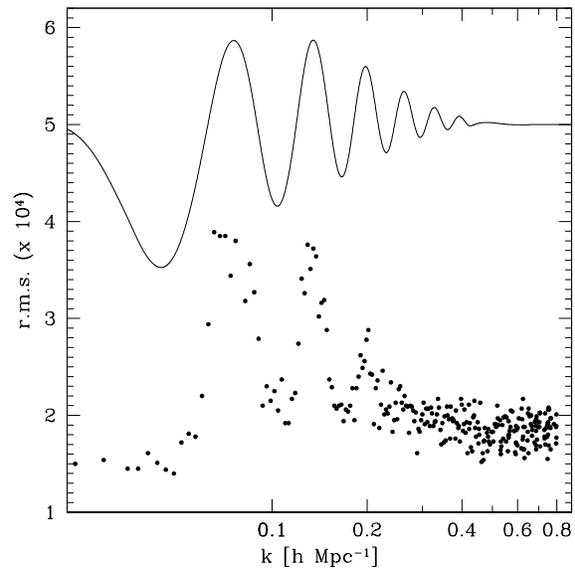}
\end{tabular}
\caption{\label{fig:modec} We show the level of uncertainty due to mode-coupling as measured for 100 pairs of PATCHY simulations with and without BAO wiggles, but sharing the same white noise. To indicate the correlation between the maxima of the mode-coupling and the BAO peaks we show the initial $P/P_{nw}$ from CAMB, scaled with an appropriate factor for visualization puroposes (thin line)}.
\end{figure}

There are two main contributions to the uncertainty in the BAO shift estimates:

i) Random errors in estimates of the power spectrum.  There is a finite number of independent harmonics conributing to the power in each bin in $k-$space used to estimate the power spectrum. The amplitude of each harmonic is a random number with gaussian distribution. The finite number of the harmonics results in a random error in the estimate of $P(k)$.

ii) Nonlinear mode-coupling may result in additional errors. The simulation including BAO wiggles will experience larger gravitational interactions at the scale of the BAO leading to correlated errors in the estimates of $P(k)$.

In order to estimate the importance of mode-coupling errors we use the PATCHY code \citep{Kitaura1,Kitaura2} to generate a large number of non-linear density field realizations. The PATCHY code uses Lagrangian perturbation theory (LPT) and a non-linear, scale-dependent stochastic biasing scheme to produce halo realisations of the density field. In particular it uses augmented LPT (ALPT, \citet{KH}), to generate a dark matter density field on a mesh starting from Gaussian fluctuations and to compute the peculiar velocity field. ALPT is based on a combination of second order LPT (2LPT) on large scales and the spherical collapse model on smaller scales. PATCHY also accounts for the missing power of perturbative approaches w.r.t. $N$-body simulations.

For our estimates we generated 100 pairs of PATCHY matter density fileds at $z = 0$ with and without BAO wiggles, but sharing the same white noise. As one can see in Figure~\ref{fig:modec} the dispersion peaks, as expected, at the BAO positions (as a reference, we also plot the mean of the $P/P_{nw}$ ratio for the 100 realisations, solid line). However, the level of uncertainty is smaller than the dispersion in the measurement of the $P/P_{nw}$ ratio in our BigMultiDark simulation, shown in Figure~\ref{fig:power}, i.e. $\sim 5 \times 10^{-4}$. The residual noise at large k's comes from power from smaller k's at the BAO scale as shown in \citet{Neyrinck} (see upper panel in their Figure 3). 

Therefore, to be conservative, we adopt for the Chi-square fitting the error estimated from the dispersion in the wavenumber range $0.5  < k < 0.8$ h Mpc$^{-1}$ that is free of oscillation features, and as mentioned above, has a level of uncertainty, $\sim 5 \times 10^{-4}$, larger than the effects due to mode-coupling, i.e. $< 4 \times 10^{-4}$.

As mentioned above, the dilation (shift) parameter $\alpha$ yields the relative position of the acoustic scale in our Planck simulations w.r.t. the model adopted in Eq.~\ref{eq:pkratio}. From our fit we measure for the dark matter tracer at $z=0$  a small BAO shift $\alpha-1 [\%]=0.353^{+0.027}_{-0.026}$ for the data shown in Figure~\ref{fig:power}. This indicates a shift of the the acoustic scale towards larger k, relative to the linear power spectrum, which it has been measured with at least 5 times better precision as previously reported in other works  \citep[e.g.][]{CroSco2008,Sanchez2008,Nikhil2009,Seo2010}. For the data of the halo biased tracer sample shown in Figure~\ref{fig:power} we measure a shift  $\alpha-1 [\%]=0.236^{+0.086}_{-0.091}$, and a damping of the BAO feature $\Sigma_{nl}=7.741^{+0.092}_{-0.088}$ smaller than that measured for dark matter $\Sigma_{nl}=8.231^{+0.025}_{-0.027}$. The bottom panels show the likelihood $\chi^2$ distributions for the BAO shift $\alpha$ parameter both for dark matter and the halo sample. 

In the next section we provide the main results of our analysis for the nonlinear evolution with redshift of the shift and damping of the BAO feature for dark matter and four halo samples with different number density, both in real- and redshift-space.

\section{Results on BAO systematics} 
\label{sec:results}

\subsection{Matter} 

\begin{figure}
\begin{tabular}{cc}
\includegraphics[width=8.5cm]{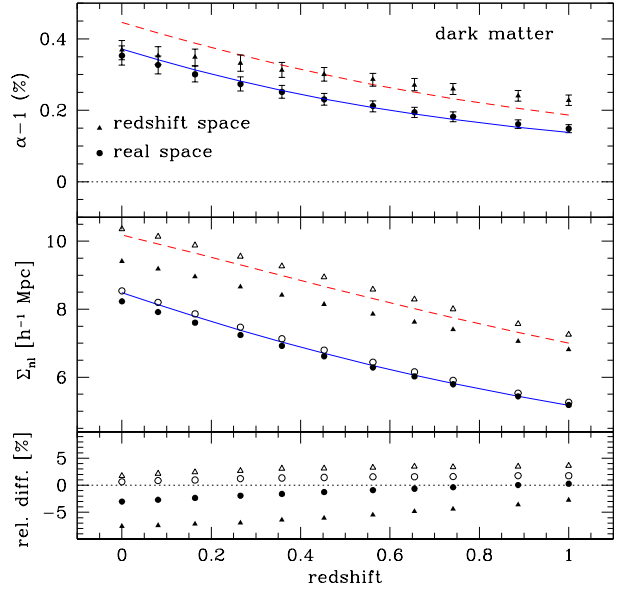}
\end{tabular}
\caption{Nonlinear evolution of the BAO shift and damping with redshift for dark matter in real- and redshift-space (solid circles and triangles respectively). The solid line in the top and middle panel are our best fit to $\alpha(z)-1 \propto [D(z)/D(0)]^2$ and the linear theory estimate of the damping given by Eq.~\ref{eq:kstar} respectively.  Errors for the damping measurements are smaller than the size of the symbols. The dashed lines correspond to redshift-space predictions. The open circles and triangles are representing the dispersion of the dark matter pair separation at BAO scales measured from the BigMD simulation (see text). The bottom panel shows the relative ratio of the damping measurements as compared to linear theory. 
}\label{fig:fig4}
\end{figure}

We show in Table~\ref{table:tab1} and Figure~\ref{fig:fig4} our main results on the nonlinear evolution with redshift of the BAO shift $\alpha-1 [\%]$ and damping $\Sigma_{nl}$ for the dark matter tracer, in real- and redshift-space, following the methodology described in Sec.~\ref{sec:bao}. The trend of the acoustic scale shift towards $z=0$ is measured at high-precision, at least 5 times better that any other estimates found in the literature as mentioned above. This, together with the good sampling in redshift, allow us to provide an accurate parametrisation of the evolution of  $\alpha$ as a function of the linear growth factor $D(z)$. For the data in real-space we find $\alpha(z)-1 [\%] = (0.350\pm 0.014)[D(z)/D(0)]^{1.74\pm 0.14}$. The measured power index is close to the expected $D(z)^2$ prediction from perturbation theory  \citep[see][]{Nikhil2009,Zaldarriaga2012}. In this case, if we fix the power index to 2, we obtained $\alpha(z)-1 [\%] = (0.3716\pm 0.0083)[D(z)/D(0)]^2$ (solid line in the top panel of Figure~\ref{fig:fig4}). These results are consistent with \citet{Seo2010}.

Moreover, the evolution of BAO damping in real-space as a function of redshift agrees remarkably well, within $0.25\%$ at $z=1$ and $3\%$ at $z=0$, with that from linear theory shown in the middle panel of Figure~\ref{fig:fig4} as solid line (see also bottom panel for the relative ratio), where the broadening and attenuation of the BAO feature is exponential, as adopted in our model, with a scale $\Sigma^{th}_{nl}$ computed following \cite{Crocce:2005xz, Matsubara}, i.e.
\begin{equation} \label{eq:kstar}
\Sigma^{th}_{nl}=\left[\frac{1}{3\pi^2}\int P_{lin}(k)dk\right]^{1/2}.
\end{equation}
BAO damping is basically introduced by the dispersion of pair separations at BAO scales. Similar to \cite{Eisenstein:2006nj}, in our Table~\ref{table:tab1}, we also show the root-mean-square of the displacements of the dark matter particle pairs with initial separations $\sim100\mpcoh$, $\Sigma_{100}$.  The displacement is defined as the difference between the initial separation and the separation at given redshift in the radial direction (along the line connecting the pair). The dispersion of the separations are consistent with our BAO damping measurements in $1-2\%$ (see also Figure~\ref{fig:fig4}).

\begin{table}\scriptsize
\begin{center}
\begin{tabular}{crrrrrrr}
\hline
redshift  	& $\alpha-1$[\%]  & $\Sigma_{nl}$ (Mpc/h) & $\Sigma_{nl}^{th}$  & $\Sigma_{100}$ & $\chi^2/d.o.f.$ \\ \hline	
1.000 	&\ \ $  0.148^{+0.011}_{-0.011}	$	&\ \ $	5.185^{+0.015}_{-0.015}	$	&\ \ 5.171	&\ \ 5.262&\ \ 1.00		\\
0.887 	&\ \ $  0.161^{+0.012}_{-0.012}	$	&\ \ $	5.440^{+0.015}_{-0.016}	$	&\ \ 5.438 	&\ \ 5.534&\ \ 1.02	        \\
0.741	&\ \ $  0.182^{+0.013}_{-0.014}	$	&\ \ $	5.795^{+0.016}_{-0.017}	$	&\ \ 5.816	&\ \ 5.910&\ \ 0.99		\\
0.655 	&\ \ $  0.194^{+0.014}_{-0.014}	$	&\ \ $        6.024^{+0.017}_{-0.017}	$	&\ \ 6.062	&\ \ 6.158&\ \ 1.03		\\
0.562 	&\ \ $  0.212^{+0.014}_{-0.016}	$	&\ \ $	6.288^{+0.018}_{-0.017}	$	&\ \ 6.345	&\ \ 6.443&\ \ 1.05		\\
0.453 	&\ \ $  0.231^{+0.017}_{-0.016}	$	&\ \ $        6.617^{+0.019}_{-0.018}	$	&\ \ 6.703	&\ \ 6.800&\ \ 1.02		\\
0.358         &\ \ $  0.251^{+0.018}_{-0.018}	$	&\ \ $        6.923^{+0.020}_{-0.019}	$	&\ \ 7.035	&\ \ 7.130&\ \ 1.03		\\
0.265 	&\ \ $  0.273^{+0.021}_{-0.019}	$	&\ \ $	7.242^{+0.021}_{-0.021}	$	&\ \ 7.384	&\ \ 7.475&\ \ 1.01		\\
0.164 	&\ \ $  0.301^{+0.023}_{-0.021}	$	&\ \ $	7.605^{+0.023}_{-0.023}	$	&\ \ 7.787 &\ \ 7.863&\ \ 1.01		\\
0.081 	&\ \ $  0.327^{+0.025}_{-0.025}	$	&\ \ $        7.916^{+0.024}_{-0.024}	$	&\ \ 8.135	&\ \ 8.205&\ \ 1.01		\\
0.000 	&\ \ $  0.353^{+0.027}_{-0.026}	$	&\ \ $        8.231^{+0.025}_{-0.027}	$	&\ \ 8.486	&\ \ 8.541&\ \ 1.01		\\
\hline
\end{tabular}
\end{center}
\caption{
The best fit values for the BAO shift $\alpha$ and damping $\Sigma_{nl}$ measured at different redshifts from fitting the real-space power spectrum $P(k)/P_{nw}(k)$ ratio drawn from all BigMD dark matter particles. The damping computed from linear theory $\Sigma^{th}_{nl}$, given Eq.~\ref{eq:kstar}, is also given for comparison. $\Sigma_{100}$ is the dispersion of the dark matter particle pair separation at BAO scales. The $\chi^2$ per degree-of-freedom are also listed.
} \label{table:tab1}
\end{table}

\begin{table}\scriptsize
\begin{center}
\begin{tabular}{crrrrrrr}
\hline
redshift  	& $\alpha-1$[\%]  & $\Sigma_{nl}$  (Mpc/h)& $\Sigma_{100}$ & $\chi^2/d.o.f.$ \\ \hline	
1.000 	&\ \ $	0.228^{+0.015}_{-0.015}	$	&\ \ $	6.814^{+0.017}_{-0.016}	$       &\ \ 	7.257&\ \ 0.98	\\
0.887 	&\ \ $	0.241^{+0.015}_{-0.015}	$	&\ \ $        7.056^{+0.016}_{-0.017}	$	&\ \ 	7.572&\ \ 1.04	\\
0.741	&\ \ $	0.256^{+0.015}_{-0.016}	$	&\ \ $        7.406^{+0.017}_{-0.016}	$	&\ \ 	8.010&\ \ 1.14	\\
0.655 	&\ \ $	0.271^{+0.018}_{-0.016}	$	&\ \ $	7.626^{+0.018}_{-0.017}	$	&\ \ 	8.291&\ \ 1.06	\\
0.562 	&\ \ $	0.287^{+0.017}_{-0.019}	$	&\ \ $        7.858^{+0.017}_{-0.018}	$	&\ \ 	8.584&\ \ 1.06	\\
0.453 	&\ \ $	0.301^{+0.019}_{-0.019}	$	&\ \ $	8.147^{+0.019}_{-0.019}	$	&\ \ 	8.942&\ \ 1.02	\\
0.358         &\ \ $	0.313^{+0.021}_{-0.020}	$	&\ \ $	8.415^{+0.018}_{-0.021}	$	&\ \ 	9.268&\ \ 1.06	\\
0.265 	&\ \ $        0.332^{+0.023}_{-0.023}	$	&\ \ $	8.657^{+0.021}_{-0.021}	$	&\ \ 	9.553&\ \ 1.02	\\
0.164 	&\ \ $	0.349^{+0.023}_{-0.024}	$	&\ \ $	8.959^{+0.021}_{-0.023}	$	&\ \ 	9.881&\ \ 1.08	\\
0.081 	&\ \ $	0.353^{+0.026}_{-0.025}	$	&\ \ $        9.185^{+0.024}_{-0.022}	$	&\ \ 	10.135&\ \ 1.02	\\
0.000 	&\ \ $        0.369^{+0.026}_{-0.027}	$	&\ \ $	9.410^{+0.024}_{-0.025}	$	&\ \ 	10.358&\ \ 1.01	\\
\hline
\end{tabular}
\end{center}
\caption{
As in Table~\ref{table:tab1} but for redshift-space.
} \label{table:tab2}
\end{table}

\begin{figure}
\begin{tabular}{cc}
\includegraphics[width=8.5cm]{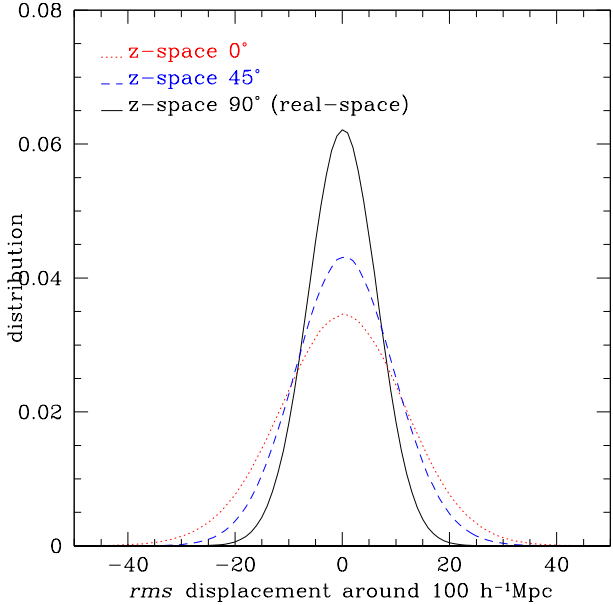}
\end{tabular}
\caption{
The root-mean-square of the displacements with initial separation of $\sim100\mpcoh$ in redshift space. We select the pairs in three different directions, 0, 45, and 90 degree with respect to the line of sight, of which the one in 90 degree (in the transverse direction) is the same as the displacement in real space since we only are interested in the radial direction (along the line connecting the pairs). One can see that the dispersion is larger when the angle between the pair and line of sight is smaller. The dispersion of the separations will erase the clustering signal and result in BAO damping.
}
\label{fig:displacement}
\end{figure}

In Table~\ref{table:tab2} and Figure~\ref{fig:fig4} we also provide the shift and damping measurements in redshift-space. In agreement with \citet[][]{Seo2010} we measure BAO shifts in redshift-space which are larger than in real-space. An increase of the shift is expected due to redshift-space distortions (RSD) induced by peculiar velocities. In linear perturbation theory of gravitational instability the Lagrangian displacement in the Zeldovich approximation is larger along the line-of-sight direction by a factor $(1+f)$, where $f = d \, \ln \, D/d\, \ln \,a \approx \Omega_m^{0.55}(z)$ is the logarithmic derivative of the linear growth rate \citep[][]{Eisenstein:2006nj}. Hence, for a spherically averaged power spectrum in redshift-space we expect a shift increase of $\approx [((1+f)^2+2)/3]^{1/2}$ (being $\sim 1+f/3$ also a good approximation as adopted in \citet[][]{Seo2010}). The dashed line in the top panel of Figure~\ref{fig:fig4} shows the theoretical prediction for the shift of the acoustic scale in redshift-space adopting our best fit to $\alpha(z)-1 \propto [D(z)/D(0)]^2$  for the shift evolution in real-space (solid line). The simulation and model shift results agree well within $\sim 20 \%$  over the entire redshift range. Following the same argument in linear theory, there should be also an increase of the BAO damping in redshift-space by the same $f$-factor as compared to that in real-space. Indeed, this is what we find as shown in the middle panel of Figure~\ref{fig:fig4}. The dashed line shows the redshift-space linear prediction for the BAO damping after we adopt Eq.~\ref{eq:kstar} for the real-space theory estimate. The agreement with our simulation results is better than $\sim 10 \%$  over the entire redshift range (see bottom panel of Figure~\ref{fig:fig4}), with an increasing departure towards $z=0$ due to nonlinear effects. In Table~\ref{table:tab2} and Figure~\ref{fig:fig4}, we also show the root-mean-square of the displacements of the dark matter particle pairs with initial separations $\sim100\mpcoh$ in redshift space. The agreement is not as good as in real-space. It might indicate that the spherically-averaged BAO damping cannot not be derived perfectly from the spherically-averaged $rms$ of the displacement because of anisotropy. This is shown in Figure~\ref{fig:displacement}) where we plot the $rms$ of the displacements with initial separation of $\sim100\mpcoh$ in redshift space. We select the pairs in three different directions, 0, 45, and 90 degree with respect to the line of sight, of which the one in 90 degree - in the transverse direction - is the same as the displacement in real space since we only are interested in the radial direction (along the line connecting the pairs). One can see that the dispersion is larger when the angle between the pair and line of sight is smaller. The dispersion of the separations will erase the clustering signal and result in BAO damping.

\subsection{Halos} 

\begin{table*}
\begin{tabular}{crrrrrr}\hline
$n=2.0 \times 10^{-3}$ & $z$=0 & 0.164  & 0.358 & 0.562 & 0.741 & 1.000\\ \hline
bias                        &\ \  1.19		&\ \ 	1.29	  &\ \ 	1.42	  &\ \ 	1.56	&\ \ 1.70 &\ \ 1.91 \\
$\alpha-1$[\%]       &\ \  $0.241^{+0.066}_{-0.067} $ &\ \ 	$0.165^{+0.074}_{-0.074} $	  &\ \ 	$0.123^{+0.060}_{-0.052} $	   &\ \ 	$0.087^{+0.060}_{-0.063} $	&\ \ 	$0.318^{+0.070}_{-0.069} $ &\ \ 	$0.356^{+0.062}_{-0.063} $ \\
$\Sigma_{nl}$        &\ \  $7.757^{+0.065}_{-0.068} $ &\ \ 	$7.048^{+0.078}_{-0.075} $   &\ \ $6.513^{+0.063}_{-0.061} $    &\ \        $6.284^{+0.068}_{-0.072} $ &\ \ 	$5.732^{+0.078}_{-0.079} $ &\ \ 	$5.335^{+0.074}_{-0.074} $ \\
$\chi^2/dof$           &\ \  1.32		&\ \ 	1.11         &\ \   1.40         &\ \  1.13  &\ \  1.38  &\ \ 1.48 \\
\hline
$n=1.0 \times 10^{-3}$\\
\hline
bias                        &\ \  1.33		&\ \ 	1.44	  &\ \ 	1.59	  &\ \ 	1.76	&\ \ 1.92 &\ \ 2.16 \\
$\alpha-1$[\%]       &\ \ $0.236^{+0.086}_{-0.091} $ &\ \  $0.083^{+0.074}_{-0.086} $	&\ \ 	$0.162^{+0.066}_{-0.070} $	  &\ \ 	$0.217^{+0.076}_{-0.081} $	&\ \ 	$0.266^{+0.076}_{-0.073} $ &\ \ 	$0.333^{+0.065}_{-0.065} $\\
$\Sigma_{nl}$        &\ \ $7.741^{+0.092}_{-0.087} $ &\ \  $6.834^{+0.082}_{-0.078} $       &\ \  $6.427^{+0.075}_{-0.075} $    &\ \ $6.289^{+0.090}_{-0.087} $  &\ \ 	$5.994^{+0.090}_{-0.085} $ &\ \ 	$5.340^{+0.084}_{-0.080} $     \\
$\chi^2/dof$           &\ \ 0.99		&\ \ 1.25	         &\ \  1.22           &\ \   0.99  &\ \ 1.03 &\ \ 1.30    \\
\hline
$n=4.0 \times 10^{-4}$\\
\hline
bias                        &\ \  1.54		&\ \ 	1.67	  &\ \ 	1.84	  &\ \ 	2.04 &\ \ 2.22 &\ \ 2.50	\\
$\alpha-1$[\%]       &\ \ $0.379^{+0.106}_{-0.103} $  &\ \ 	$0.133^{+0.099}_{-0.114} $	  &\ \ 	$0.231^{+0.098}_{-0.094} $	  &\ \ 	$0.377^{+0.098}_{-0.095} $	&\ \ 	$0.277^{+0.103}_{-0.089} $ &\ \ 	$0.25^{+0.096}_{-0.085} $\\
$\Sigma_{nl}$        &\ \ $7.430^{+0.120}_{-0.103} $  &\ \ 	$6.728^{+0.106}_{-0.110} $   &\ \  $6.408^{+0.108}_{-0.098} $ &\ \   $6.081^{+0.120}_{-0.115} $ &\ \ 	$5.542^{+0.124}_{-0.108} $ &\ \ 	$5.182^{+0.114}_{-0.107} $     \\
$\chi^2/dof$           &\ \ 1.25		&\ \ 1.18	         &\ \  1.28           &\ \   1.02  &\ \ 0.97 &\ \ 1.12    \\
\hline
$n=2.0 \times 10^{-4}$\\
\hline
bias                        &\ \ 1.73		&\ \ 	1.88	  &\ \ 	2.07	  &\ \ 	2.28	&\ \ 2.49 &\ \ 2.80 \\
$\alpha-1$[\%]      &\ \ $0.451	^{+0.162}_{-0.170} $ &\ \ 	$0.127^{+0.150}_{-0.165} $	  &\ \ 	$0.434^{+0.129}_{-0.147} $	  &\ \ 	$0.265^{+0.135}_{-0.131} $	&\ \ 	$0.350^{+0.118}_{-0.122} $ &\ \ 	$0.111^{+0.122}_{-0.117} $\\
$\Sigma_{nl}$        &\ \ $7.550^{+0.171}_{-0.171} $  &\ \ 	$6.557^{+0.168}_{-0.157} $  &\ \  $6.015^{+0.136}_{-0.136} $  &\ \  $5.931^{+0.162}_{-0.171} $  &\ \ 	$5.159^{+0.144}_{-0.130} $ &\ \ 	$4.928^{+0.137}_{-0.147} $     \\
$\chi^2/dof$           &\ \ 0.83		&\ \ 	0.88         &\ \ 1.02            &\ \    0.88 &\ \  1.09 &\ \ 1.03   \\

\hline
\end{tabular}
\caption{
The best fit values for the BAO shift $\alpha$ and damping $\Sigma_{nl}$ measured at several redshifts from fitting the $P(k)/P_{nw}(k)$ ratio, in real-space, drawn from four different number densities (and bias) BigMD halo samples. The $\chi^2$ per degree-of-freedom are also provided.
} \label{table:tab3}
\end{table*}

\begin{table*}
\begin{tabular}{crrrrrr}\hline
$n=2.0 \times 10^{-3}$ & $z$=0 & 0.164  & 0.358 & 0.562 & 0.741 & 1.000\\ \hline
$\alpha-1$[\%]       &\ \ $0.264^{+0.124}_{-0.113} $  &\ \ $0.224^{+0.093}_{-0.097} $  &\ \  $0.212^{+0.087}_{-0.097} $	  &\ \ 	$0.140^{+0.079}_{-0.089} $	  &\ \  $0.403^{+ 0.074}_{-0.071} $ 	 &\ \ $0.435^{+0.063}_{-0.063} $ 	 \\
$\Sigma_{nl}$        &\ \ $9.270^{+0.105}_{-0.097} $  &\ \  $8.733^{+0.088}_{-0.086} $  &\ \  $8.092^{+0.081}_{-0.084} $       &\ \ $7.935^{+0.078}_{-0.080} $  &\ \ $7.522^{+ 0.071}_{-0.075} $  &\ \ $7.093^{+ 0.064}_{-0.067} $	 \\
$\chi^2/dof$           &\ \ 0.92	&\ \ 	1.09         &\ \   1.32          &\ \  1.04  &\ \ 1.15  &\ \ 1.53      \\
\hline
$n=1.0 \times 10^{-3}$\\
\hline
$\alpha-1$[\%]       &\ \ $0.358^{+0.134}_{-0.136} $ &\ \ $0.167^{+ 0.126}_{-0.133} $  &\ \  $0.281^{+0.120}_{-0.111} $	  &\ \ 	  $0.287^{+0.110}_{-0.116} $  &\ \ $0.391^{+0.095}_{-0.098} $ &\ \  $0.473^{+0.093}_{-0.083} $  \\
$\Sigma_{nl}$        &\ \ $9.312^{+0.129}_{-0.115} $ &\ \  $8.547^{+ 0.126}_{-0.120} $ &\ \   $8.108^{+0.107}_{-0.107} $         &\ \   $7.932^{+0.108}_{-0.109} $  &\ \ $7.720^{+0.099}_{-0.093} $ &\ \  $7.064^{+0.093}_{-0.090} $  \\
$\chi^2/dof$           &\ \ 0.82		&\ \ 	 0.99        &\ \  0.86     &\ \   0.84 &\ \ 0.83 &\ \  1.08  \\
\hline
$n=4.0 \times 10^{-4}$\\
\hline
$\alpha-1$[\%]       &\ \ $0.409^{+0.174}_{-0.157} $	  	&\ \ 	$0.217^{+0.165 }_{-0.173} $	  &\ \ 	$0.258^{+0.170}_{-0.170} $	  &\ \ 	$0.399^{+0.166}_{-0.162} $ &\ \ $0.323^{+0.126}_{-0.125} $ &\ \ $0.448^{+0.118}_{-0.117} $ 	  \\
$\Sigma_{nl}$        &\ \ $9.101^{+0.149}_{-0.147} $		&\ \ 	$8.325^{+0.147}_{-0.146} $          &\ \  $8.066^{+0.156}_{-0.157} $ &\ \ $7.766^{+0.163}_{-0.171} $ &\ \ $7.310^{+0.131}_{-0.125} $ &\ \   $7.030^{+0.128}_{-0.122} $    \\
$\chi^2/dof$           &\ \ 0.91		&\ \ 	  1.06       &\ \   0.874         &\ \ 0.67  &\ \ 0.91 &\ \  0.90    \\
\hline
$n=2.0 \times 10^{-4}$\\
\hline
$\alpha-1$[\%]       &\ \ $0.630^{+ 0.304}_{-0.309} $	  	&\ \ 	$0.127^{+0.255}_{-0.260} $	  &\ \ 	$0.434^{+0.264}_{-0.242} $	  &\ \ 	$0.402^{+0.235}_{-0.233} $ &\ \ $0.358^{+0.198}_{-0.208} $&\ \	$0.185^{+ 0.165}_{-0.171} $  \\
$\Sigma_{nl}$        &\ \ $9.403^{+ 0.275}_{-0.256} $		&\ \ 	 $8.094^{+0.234}_{-0.246} $         &\ \  $7.753^{+0.213}_{-0.226} $          &\ \ $7.591^{+ 0.243}_{-0.244} $  &\ \ $6.983^{+0.205}_{-0.199} $ &\ \   $6.782^{+0.179}_{-0.185} $      \\
$\chi^2/dof$           &\ \  0.63		&\ \ 	0.80         &\ \   0.82          &\ \ 0.60  &\ \ 0.72 &\ \ 0.79     \\
\hline
\end{tabular}
\caption{
Same as Table~\ref{table:tab3} but for redshift-space.
} \label{table:tab4}
\end{table*}

\begin{figure}
\begin{tabular}{cc}
\includegraphics[width=8.5cm]{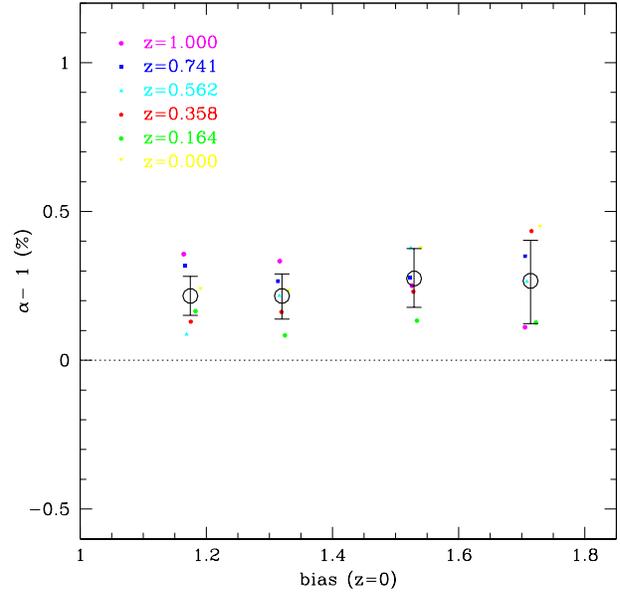}
\end{tabular}
\caption{\label{fig:fig5}  Measurements of the BAO shift as a function of halo bias for our BigMD Planck data. Each of the individual shift estimates are shown with tiny solid symbols for the different redshifts (color coded). Large open circles shows the mean values for the four bias bins, and $1\sigma$ error bars correspond to the errors of the mean.}
\end{figure}

\begin{figure}
\begin{tabular}{cc}
\includegraphics[width=8.5cm]{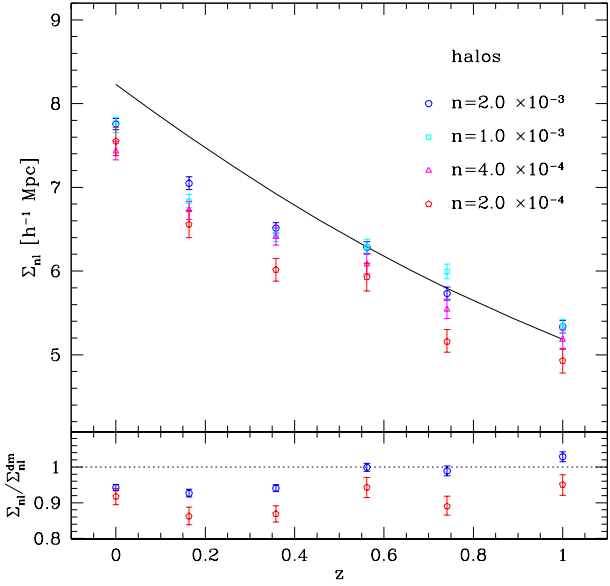}
\end{tabular}
\caption{\label{fig:fig6} Nonlinear evolution of the BAO damping with redshift, in real space, for our four different halo samples (open symbols). The solid line connects the measurements for the dark matter tracer provided in Table~\ref{table:tab1} and shown in Figure~\ref{fig:fig4}. The acoustic feature suffers less damping due to nonlinear effects as compare to dark matter towards lower redshift and also for more sparse halo samples at a given redshift.}
\end{figure}

\begin{figure}
\begin{tabular}{cc}
\includegraphics[width=8.5cm]{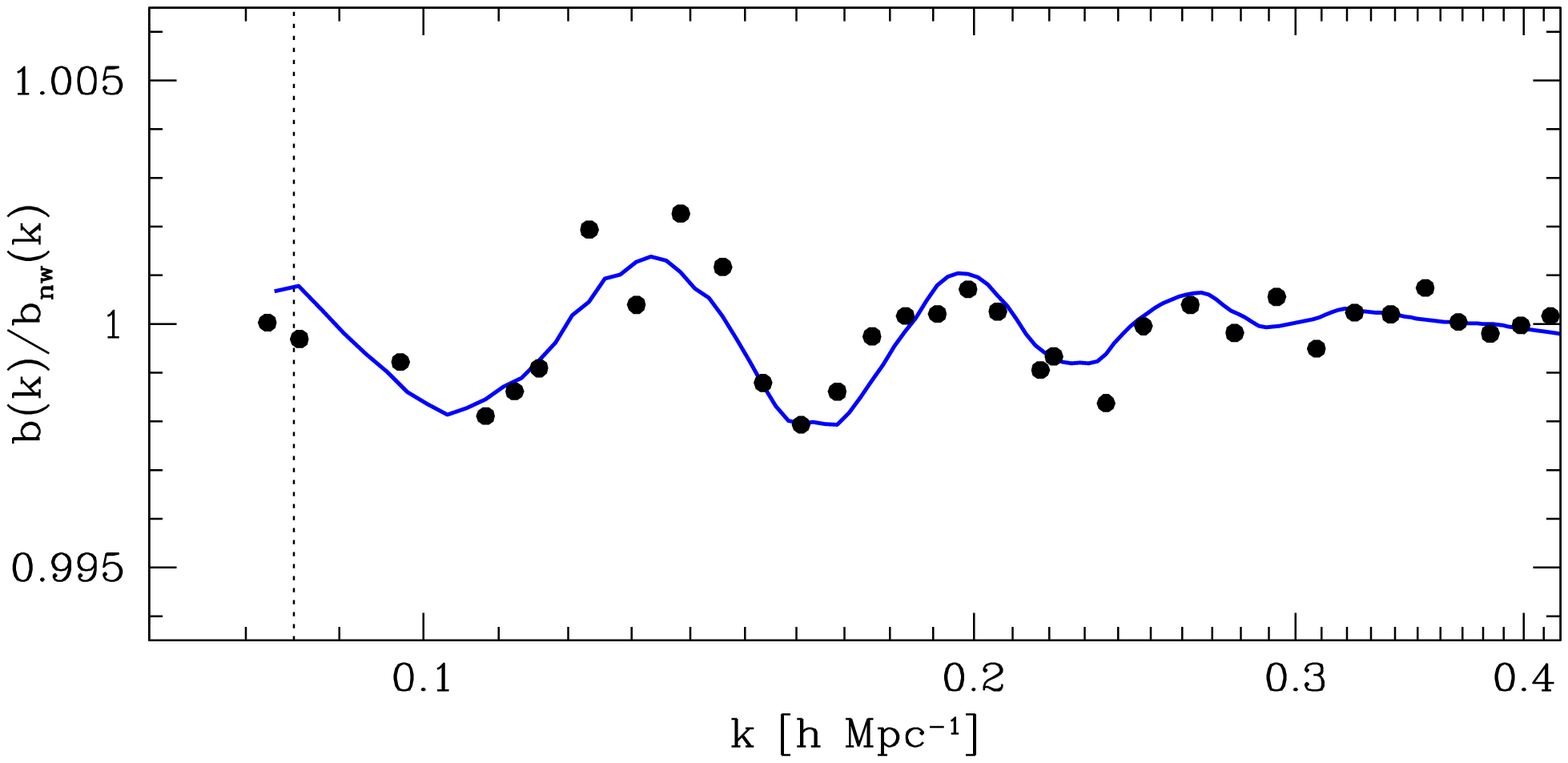}
\end{tabular}
\caption{\label{fig:fig7}  Ratio of the halo bias at $z=0$ in the BigMD Planck simulation to that in the non-wiggle realization for the halo sample with $n=1$ $\times$ 10$^{-3}$ Mpc$^{-3}$ $h^3$ and bias $1.33$(solid symbols). Our best fit model is also shown with a solid line.}
\end{figure}

We provide in Table~\ref{table:tab3} the best fit values of the BAO shift and damping measured at different redshifts up to $z=1$ for four BigMD dark matter halo samples, in real-space, with number densities $2\times10^{-3}, 1\times10^{-3}, 4\times10^{-4}, 2\times10^{-4} \Mpcvol$; which scan a halo bias range from $1.2$ to $2.8$ over the entire explored redshift range. We measure in all halo samples a typical BAO shift of $\approx 0.25 \%$ which does not seems to evolve with redshift within the errors, that range from $\sim 0.065\%$ for the denser halo sample with number density $2\times10^{-3} \Mpcvol$ to $\sim 0.14\%$ in the case of the sparsest sample with $n = 2\times10^{-4} \Mpcvol$. We summarize our BAO shift results for the halo tracers in Figure~\ref{fig:fig5}. We plot the shift $\alpha-1 [\%]$ as a function of halo bias at $z=0$, where we have adopted the assumption that $b(z=0)\equiv b(z) D(z=0)/D(z)$ \citep[][]{Lahav}. For each halo sample, we plot with different tiny solid symbols for the different redshifts (color coded) each of the individual BAO shift measurements. Additionally, large open circles shows the mean values of the shift measurements for the four bias bins, and $1\sigma$ error bars correspond to the errors of the mean. We obtain $\alpha-1 [\%]$ shifts $0.216\pm0.065, 0.216\pm0.076, 0.275\pm0.099$ and $0.267\pm0.140$ for a mean halo bias of $1.17, 1.32, 1.53$, and $1.71$ respectively. The size of the bias bins are about the diameter of the open symbols. Thus, we conclude there is a flat dependence of the BAO shift as a function of halo bias in our BigMD Planck simulations. This is the first time that this level of accuracy has been reached for the study of the shift in the acoustic scale for halo tracers. Most of works found in the literature adopted a Halo Occupation Distribution (HOD) modelling given their hardness to resolve well halos and subhalos for the bias samples studied in this work, which are typical tracers of large galaxy surveys. This result has to be reconciled with perturbation theory predictions found in the literature that predicted an increase of the shift with halo bias \citep[][]{Nikhil2009}. A better understanding of the non-local bias would be required to allow the proper shifts of the acoustic scale reported in this work for the halo tracers. Understanding this BAO systematics is key, and represent a serious challenge for future redshift surveys, such as DESI and Euclid, that aim to reach an accuracy in the BAO scale better than $\sim 0.3\%$. Redshift-space estimates for the BAO shifts are provided in Table~\ref{table:tab4}. Results, in general, may suggest a larger shift as compared to real-space, although they are not as convincing given their larger errors, about $\sim 1.5-2$ times larger than real-space uncertainties (see Table~\ref{table:tab3}).

Our measurement results in real-space on the evolution with redshift of the BAO damping for our four BigMD halo samples are also given in Table~\ref{table:tab3}. We report a different behaviour as that seen, and discussed above, for the dark matter tracer. This is clearly observe in Figure~\ref{fig:fig6}, where we see that the damping $\Sigma_{nl}$ measurements in halos decreases down to $\sim 10-15\%$ with decreasing redshift as compared to dark matter, represented in the plot by a solid line that connects the measurements listed in Table~\ref{table:tab1}, and shown in Figure~\ref{fig:fig4} as open triangles. We also notice, at a given redshift, less damping of the BAO signature for halo samples with smaller number densities, i.e. larger halo bias (see bottom panel of Figure~\ref{fig:fig6}). In summary, from our analysis we can conclude that the acoustic feature for halos suffers less broadening due to nonlinear effects than for dark matter, and remarkably, damping of the acoustic oscillations in halos seems to depends mildly on bias.

The smaller damping seen in the BAO signal for halo tracers is highlighted in Figure~\ref{fig:fig7} when we show the ratio of the scale-dependent halo bias in the BigMDPL simulation with acoustic oscillations to that in the non-wiggle BigMDPLnw simulation for the measurements of the $z=0$ halo sample with number density $1 \times$ 10$^{-3}$ Mpc$^{-3}$ $h^3$ ($bias=1.33$) shown in Figure~\ref{fig:power}, i.e. $b(k)/b_{nw}(k)\equiv[P^h(k)/P^{dm}(k)]/[P^h_{nw}(k)/P^{dm}_{nw}(k)$] (solid circles). The solid line represents the best fit model to the data as discussed in Sec. 3. A clear modulation in phase with the acoustic scale is observed in the halo bias,  with an amplitude of $\sim 0.25\%$,  due to the presence of the baryonic acoustic oscillations. It is worth mentioning the work by \citet[][]{WangZhan}, who using much worse numerical resolution and significantly less volume in their simulations, but performing many realisations and taking advantage of using non-wiggle realisations, did also detect this signature in the halo bias due to the BAO for a halo sample with two order of magnitude $\sim 10^{-3}$ Mpc$^{-3}$ h$^3$ higher number density, where they detected a modulation amplitude of $\sim 0.5\%$ due to the much larger bias of their sample ($b>3$) (see bottom panel of their Fig. 4). 

For all halo samples and redshifts we clearly observe a larger BAO damping in redshift-space. Our measurements are provided in Table~\ref{table:tab4}, and within the errors they can be explained well by adopting a shift increase of $\approx [((1+f)^2+2)/3]^{1/2}$ due to redshift-space distortions, as discussed above.
 
\section{Summary} 
\label{sec:conclusion}

We study, from redshift 1 to the present, the nonlinear evolution
of the shift and damping of baryonic acoustic oscillations in the
power spectrum of dark matter and halo tracers using cosmological
simulations with high-mass resolution over a large volume. The results presented in this paper are based on the BigMultiDark suite of simulations in the standard $\Lambda$CDM cosmology, with numerical parameters (mass and force resolution) that have been chosen to face the requirements imposed by current and future dark energy experiments on understanding any possible systematic shifts in the BAO signal due to nonlinear gravitational growth, scale-dependent bias and redshift space distortions to a high precision, better than the measured statistical uncertainties. Our measurements can also be useful for comparison with perturbation theory works that aim at explaining the nature of BAO shift and damping in dark matter and haloes.

Our main results can be summarised as follows.

\begin{itemize}

\item[(i)] For a given tracer, we measure at several redshifts the BAO shift $\alpha$ and damping $\Sigma_{nl}$ by fitting the ratio $P/P_{nw}$ of the power spectrum with acoustic oscillations to that with non-wiggle drawn from the Planck BigMDPL/BigMDPLnw simulation pair, adopting the model given in Eq.~\ref{eq:pkratio}. The effect of cosmic variance is largely reduced when dividing by the no-BAO power spectrum, which together with the proper numerical resolution and large volume allow us to report measurements of the BAO shift and damping
with unprecedented accuracy.

\item[(ii)] For dark matter, we report shifts of the acoustic scale towards larger k, relative to the linear power spectrum, measured with much better precision as previously reported in the literature. This, together with the good sampling in redshift allow us to provide an accurate parametrisation of the evolution of $\alpha$ as a function of the linear growth factor $D(z)$, i.e. $\alpha(z)-1 [\%] = (0.350\pm 0.014)[D(z)/D(0)]^{1.74\pm 0.14}$ for the data in real-space. And we find $\alpha(z)-1 [\%] = (0.3716\pm 0.0083)[D(z)/D(0)]^2$  if we fix the power index to 2, as expected from perturbation theory. Furthermore, the evolution of BAO damping $\Sigma_{nl}$ in real-space agrees remarkably well with that from linear theory as given by Eq.~\ref{eq:kstar}. In redshift-space, we measure an increase of the shift and damping as compared to real-space, which is well described in linear theory by a constant factor that depends on $f$, the logarithmic derivative of the linear growth rate.  

\item[(iii)] We measure BAO shift and damping also for four halo samples with number densities that scan a halo bias ranging from $1.2$ to $2.8$ over the entire explored redshift range.  Our BigMD simulations allows to resolve well halos and subhalos in those samples, contrary to previous studies that adopted an HOD modelling due to their lack of numerical resolution. We measure in all halo samples a typical BAO shift of $\approx 0.25\%$ in real-space, which does not seem to evolve with redshift within the uncertainties. Moreover, we report a constant shift as a function of halo bias. Redshift-space measurements are also performed, although the larger errors prevent us from a conclusive larger shift as compare to real-space. The damping of the acoustic feature for all halo samples shows a different behaviour as compared to that for dark matter. In summary, we see that halos suffer less damping, with some weak dependence on bias. A larger BAO damping is measured in redshift-space, which can be well explained by an increase of a $f$-dependent factor in linear theory due to redshift-space distortions.

\item[(iv)] A clear modulation in phase with the acoustic scale is  observed in the scale-dependent halo bias due to the presence of the baryonic acoustic oscillations when we study the ratio of the scale-dependent bias in the BigMD simulation with BAO to that in the "BAO-free" simulation. This result motivates a better understanding of non-local bias.

\end{itemize}

\section*{Acknowledgements}
We would like to thank Martin Crocce, Daniel Eisenstein, Nikhil Padmanabhan, Uros Seljak, and Hee-Jong Seo for useful discussions.
C.C. and F.P. acknowledge support from the Spanish MICINNs Consolider-Ingenio 2010 Programme under grant MultiDark CSD2009-00064, AYA2010-21231-C02-01, and MINECO Centro de Excelencia Severo Ochoa Programme under grant SEV-2012-0249. CGS acknowledges funding from the Spanish MINECO under the project AYA2012-3972-C02-01. GY thanks support from the Spanish MINECO under research grants AYA2012-31101 and  FPA2012-34694. CZ acknowledges support from Charling Tao and her grant from Tsinghua
University, and 973 program No. 2013CB834906. CZ also thanks the support from the MultiDark summer student program to visit the Instituto de F\'{\i}sica Te\'orica UAM/CSIC. The BigMultidark simulations  have been performed on the SuperMUC supercomputer  at the Leibniz-Rechenzentrum (LRZ) in Munich,  using the computing resources  awarded to  the PRACE  project  number 2012060963.

\label{lastpage}


\begin{thebibliography}{99}

\bibitem[\protect\citeauthoryear{Alimi et al.}{2012}]{Alimi2012} 
Alimi J.-M., et al., 2012, arXiv:1206.2838

\bibitem[\protect\citeauthoryear{Anderson et 
al.}{2014}]{Anderson2014} Anderson L., et al., 2014, \mnras, 441, 24
  
\bibitem[\protect\citeauthoryear{Angulo et al.}{2008}]{Angulo2008} 
Angulo R.~E., Baugh C.~M., Frenk C.~S., Lacey C.~G., 2008, \mnras, 383, 755

\bibitem[\protect\citeauthoryear{Angulo et al.}{2012}]{Angulo2012} 
Angulo R.~E., Springel V., White S.~D.~M., Jenkins A., Baugh C.~M., Frenk 
C.~S., 2012, \mnras, 426, 2046

\bibitem[\protect\citeauthoryear{Angulo et al.}{2013}]{Angulo2013} 
Angulo R.~E., White S.~D.~M., Springel V., Henriques B., 2013, 
arXiv:1311.7100

\bibitem[Blake et al.(2011)]{Blake:2011wn} 
  Blake, C., et al., 2011,
  \mnras, 415, 2892
  
\bibitem[Chuang, Wang, \& Hemantha(2012)]{Chuang:2012dv}
  Chuang, C.~-H.; Wang, Y.; and Hemantha, M.~D.~P., 2012
   \mnras, 423, 1474
  
 \bibitem[\protect\citeauthoryear{Cole et al.}{2005}]{Cole2005} 
Cole S., et al., 2005, \mnras, 362, 505

\bibitem[\protect\citeauthoryear{Conroy, Wechsler, 
\& Kravtsov}{2006}]{Conroy2006} Conroy C., Wechsler R.~H., Kravtsov A.~V., 2006, ApJ, 647, 201 
  
\bibitem[Crocce \& Scoccimarro(2006)]{Crocce:2005xz}
  Crocce, M., and Scoccimarro, R., 2006
  PhRvD, 73, 063520 
 
 \bibitem[\protect\citeauthoryear{Crocce 
\& Scoccimarro}{2008}]{CroSco2008} Crocce M., Scoccimarro R., 2008, PhRvD, 77, 023533 

\bibitem[\protect\citeauthoryear{Crocce et al.}{2010}]{Crocce2010} 
Crocce M., Fosalba P., Castander F.~J., Gazta{\~n}aga E., 2010, MNRAS, 403, 
1353 

  
  
\bibitem[Eisenstein et al.(2005)]{Eisenstein:2005su}
  Eisenstein, D.~J., et al. [SDSS Collaboration], 2005
  ApJ,  633, 560 

\bibitem[Eisenstein, Seo, \& White(2007)]{Eisenstein:2006nj} 
  Eisenstein, D.~J.;~Seo, H.~-j.; and~White, M.~J., 2007
  ApJ,  664, 660
  
  
  
\bibitem[\protect\citeauthoryear{Kim et al.}{2011}]{Kim2011} 
Kim J., Park C., Rossi G., Lee S.~M., Gott J.~R., III, 2011, JKAS, 44, 217  

\bibitem[\protect\citeauthoryear{Kitaura 
\& He{\ss}}{2013}]{KH} Kitaura F.-S., He{\ss} S., 2013, MNRAS, 435, L78 

\bibitem[\protect\citeauthoryear{Kitaura, Yepes, 
\& Prada}{2014}]{Kitaura1} Kitaura F.-S., Yepes G., Prada F., 2014, MNRAS, 439, L21 

\bibitem[\protect\citeauthoryear{Kitaura et 
al.}{2014}]{Kitaura2} Kitaura F.-S., Gil-Mar{\'{\i}}n H., 
Scoccola C., Chuang C.-H., M{\"u}ller V., Yepes G., Prada F., 2014, arXiv:1407.1236 

\bibitem[\protect\citeauthoryear{Klypin et al.}{2013}]{Klypin2013} 
Klypin A., Prada F., Yepes G., Hess S., Gottlober S., 2013, arXiv, 
arXiv:1310.3740

\bibitem[\protect\citeauthoryear{Klypin et al.,}{in prep.}]{Klypin2014} 
Klypin et al., in preparation
  
\bibitem[\protect\citeauthoryear{Klypin 
\& Holtzman}{1997}]{KH1997} Klypin A., Holtzman J., 1997, arXiv:astro-ph/9712217
  
\bibitem[\protect\citeauthoryear{Lahav et al.}{2002}]{Lahav} 
Lahav O., et al., 2002, MNRAS, 333, 961

\bibitem[Laureijs et al.(2011)]{Laureijs:2011}
Laureijs, R., et al., ``Euclid Definition Study Report'', arXiv:1110.3193

\bibitem[\protect\citeauthoryear{Lawrence et 
al.}{2010}]{Lawrence2010} Lawrence E., Heitmann K., White M., Higdon 
D., Wagner C., Habib S., Williams B., 2010, ApJ, 713, 1322

\bibitem[Lewis, Challinor, \& Lasenby(2000)]{Lewis:1999bs}
  Lewis, A.; Challinor, A.; and Lasenby,
A., 2000
  ApJ,  538, 473
  
\bibitem[\protect\citeauthoryear{Levi et al.}{2013}]{Levi2013} 
Levi M., et al., 2013, arXiv:1308.0847 
  

\bibitem[Matsubara(2008)]{Matsubara}
  Matsubara, T., 2008,
  PhRevD, 77, 063530 

\bibitem[\protect\citeauthoryear{McBride et al.,}{in prep.}]{McBride} 
McBride et al., in preparation

\bibitem[\protect\citeauthoryear{Mehta et al.}{2011}]{Mehta2011} 
Mehta K.~T., Seo H.-J., Eckel J., Eisenstein D.~J., Metchnik M., Pinto P., 
Xu X., 2011, ApJ, 734, 94 

\bibitem[\protect\citeauthoryear{Neyrinck 
\& Yang}{2013}]{Neyrinck} Neyrinck M.~C., Yang L.~F., 2013, MNRAS, 433, 1628 
  
\bibitem[Nuza et al.(2012)]{Nuza:2012mw} 
  Nuza, S.~E.;~Sanchez, A.~G.;~Prada, F.;~Klypin, A.;~Schlegel, D.~J.;~Gottloeber, S., et al., 2012,
  arXiv:1202.6057

  
\bibitem[\protect\citeauthoryear{Padmanabhan 
\& White}{2009}]{Nikhil2009} Padmanabhan N., White M., 2009, PhRvD, 80, 063508

\bibitem[Padmanabhan et al.(2012)]{Padmanabhan:2012hf} 
  Padmanabhan, N., et al., 2012,
  arXiv:1202.0090
  

\bibitem[Percival et al.(2007)]{Percival:2007yw}
  Percival, W.~J., et al., 2007,
 \mnras,  381, 1053 
    
\bibitem[\protect\citeauthoryear{Planck Collaboration et 
al.}{2013}]{Planck2013} Planck Collaboration, et al., 2013,
arXiv:1303.5076

\bibitem[Press et al.(1992)]{press92}
    Press W.H., Teukolsky S,A., Vetterling W.T., Flannery B.P.,
    1992, Numerical recipes in C. The art of scientific computing, 
    Second edition, Cambridge: University Press. 


\bibitem[\protect\citeauthoryear{Prada et al.}{2012}]{Prada2012} 
Prada F., Klypin A.~A., Cuesta A.~J., Betancort-Rijo J.~E., Primack J., 
2012, \mnras, 423, 3018 

\bibitem[\protect\citeauthoryear{Rasera et al.}{2014}]{Rasera2014} 
Rasera Y., Corasaniti P.-S., Alimi J.-M., Bouillot V., Reverdy V., 
Balm{\`e}s I., 2014, MNRAS, 440, 1420 

\bibitem[Reid et al.(2010)]{Reid:2010xm} 
  Reid, B.~A.; Percival, W.~J.; Eisenstein, D.~J.; Verde, L.; Spergel, D.~N.; Skibba, R.~A.; Bahcall, N.~A.; and Budavari, T., et al., 2010,
 \mnras,  404, 60

\bibitem[\protect\citeauthoryear{Riebe et al.}{2011}]{Riebe2011} 
Riebe K., et al., 2011, arXiv:1109.0003

\bibitem[\protect\citeauthoryear{S{\'a}nchez, Baugh, 
\& Angulo}{2008}]{Sanchez2008} S{\'a}nchez A.~G., Baugh C.~M., Angulo R.~E., 2008, \mnras, 390, 1470

\bibitem[Sanchez et al.(2009)]{Sanchez:2009jq}
  Sanchez, A.~G.; Crocce, M.; Cabre, A.; Baugh, C.~M.; and Gaztanaga,
E., 2009,
 \mnras,  400, 1643
  
\bibitem[Schlegel et al.(2011)]{Schelgel:2011zz} 
  Schlegel, D. et al.  [BigBOSS Collaboration], 2011,
  arXiv:1106.1706
  
\bibitem[Seo \& Eisenstein(2003)]{SE03}
Seo, H., Eisenstein, D. J., 2003, ApJ, 598, 720

\bibitem[\protect\citeauthoryear{Seo et al.}{2008}]{Seo2008} 
Seo H.-J., Siegel E.~R., Eisenstein D.~J., White M., 2008, ApJ, 686, 13

\bibitem[\protect\citeauthoryear{Seo et al.}{2010}]{Seo2010} 
Seo H.-J., et al., 2010, ApJ, 720, 1650 

\bibitem[\protect\citeauthoryear{Sherwin 
\& Zaldarriaga}{2012}]{Zaldarriaga2012} Sherwin B.~D., Zaldarriaga M., 2012, PhRvD, 85, 103523 

  
\bibitem[Smith et al.(2003)]{Smith:2002dz}
  Smith, R.~E., {\it et al.}  [The Virgo Consortium Collaboration], 2003,
 \mnras, 341, 1311
  [arXiv:astro-ph/0207664].
  
\bibitem[\protect\citeauthoryear{Smith, Scoccimarro, 
\& Sheth}{2008}]{Smith2008} Smith R.~E., Scoccimarro R., Sheth R.~K., 2008, PhRvD, 77, 043525 

\bibitem[\protect\citeauthoryear{Smith et al.}{2012}]{Smith2012} 
Smith R.~E., Reed D.~S., Potter D., Marian L., Crocce M., Moore B., 2012, 
arXiv:1211.6434 

\bibitem[\protect\citeauthoryear{Skillman et 
al.}{2014}]{Skillman2014} Skillman S.~W., Warren M.~S., Turk M.~J., 
Wechsler R.~H., Holz D.~E., Sutter P.~M., 2014, arXiv:1407.2600 

  
  \bibitem[\protect\citeauthoryear{Teyssier et 
al.}{2009}]{Teyssier2009} Teyssier R., et al., 2009, A\&A, 497, 335

   
\bibitem[\protect\citeauthoryear{Trujillo-Gomez et 
al.}{2011}]{T2011} Trujillo-Gomez S., Klypin A., Primack J., 
Romanowsky A.~J., 2011, ApJ, 742, 16 
  


\bibitem[Wang, Chuang \& Mukherjee(2012)]{Wang:2011sb} 
  Wang, Y.; Chuang, C.~-H.; and Mukherjee, P., 2012,
  PhRvD, 85, 023517 
  
\bibitem[\protect\citeauthoryear{Wang 
\& Zhan}{2013}]{WangZhan} Wang Q., Zhan H., 2013, ApJ, 768, L27 

\bibitem[\protect\citeauthoryear{Weinberg et 
al.}{2012}]{Weinberg2012} Weinberg D.~H., Mortonson M.~J., 
Eisenstein D.~J., Hirata C., Riess A.~G., Rozo E., 2012,
arXiv:1201.2434 

   
\bibitem[Xu et al.(2012)]{Xu12}
Xu, X., et al., 2012, arXiv:1206.6732


\bibitem[\protect\citeauthoryear{Watson et al.}{2013}]{Watson2013} 
Watson W.~A., Iliev I.~T., D'Aloisio A., Knebe A., Shapiro P.~R., Yepes G., 
2013, MNRAS, 433, 1230 

   

\end{thebibliography}
\end{document}